\documentclass[aps, prb, reprint, twocolumn, superscriptaddress]{revtex4-1}

\usepackage{multirow}
\usepackage{graphicx}
\usepackage{longtable}
\usepackage[utf8]{inputenc}
\usepackage[T1]{fontenc}
\usepackage{epstopdf}
\usepackage{textcomp}
\usepackage[cmyk,dvipsnames]{xcolor} 
\usepackage{amsbsy}
\usepackage{amsmath}
\usepackage{amssymb}
\usepackage{amsfonts}
\usepackage{dsfont}
\usepackage{mathtools}
\usepackage{braket}
\usepackage{array}
\usepackage{gensymb}
\usepackage{placeins}
\usepackage{dashrule}
\usepackage{xcolor}
\usepackage{booktabs}
\usepackage{hyperref}
\usepackage{float}
\usepackage{braket}
\usepackage{bm}
\usepackage[draft]{pgf}
\usepackage{pgfplots, pgfplotstable}
\usepackage{tikz}
\usetikzlibrary{positioning}

\usepackage[normalem]{ulem}

\pgfplotsset{compat=1.14}

\makeatletter
  \def\my@tag@font{\normalsize}
  \def\maketag@@@#1{\hbox{\m@th\normalfont\my@tag@font#1}}
  \let\amsmath@eqref\eqref
  \renewcommand\eqref[1]{{\let\my@tag@font\relax\amsmath@eqref{#1}}}
\makeatother

\renewcommand{\vec}[1]{\bm{#1}}
\newcommand{\nvec}[1]{\hat{\bm{#1}}}

\allowdisplaybreaks

\begin{document}

\title{ Spirit: Multifunctional Framework for Atomistic Spin Simulations}

\newcommand{\fz}{Peter Gr\"unberg Institut and Institute for Advanced Simulation, Forschungszentrum J\"ulich and JARA, 52425 J\"ulich, Germany}
\newcommand{\iceland}{Science Institute and Faculty of Physical Sciences, University of Iceland, VR-III, 107 Reykjav\'{i}k, Iceland}
\newcommand{\rwth}{Department of Physics, RWTH Aachen University, 52056 Aachen, Germany}

\author{Gideon P. M\"uller}
 	\homepage[]{https://juspin.de}
	\email[]{g.mueller@fz-juelich.de}
 	\affiliation{\fz}
	\affiliation{\iceland}
 	\affiliation{\rwth}
\author{Markus Hoffmann}
	\affiliation{\fz}
\author{Constantin Di\ss elkamp}
    \affiliation{\fz}
	\affiliation{\rwth}
\author{Daniel Sch\"urhoff}
    \affiliation{\fz}
	\affiliation{\rwth}
\author{Stefanos Mavros}
    \affiliation{\fz}
	\affiliation{\rwth}
\author{Moritz Sallermann}
	\affiliation{\fz}
\author{Nikolai S. Kiselev}
	\affiliation{\fz}
\author{Hannes J\'onsson}
	\affiliation{\iceland}
\author{Stefan Bl\"ugel}
	\affiliation{\fz}

\date{\today}

\begin{abstract}
The \textit{Spirit} framework is designed for atomic scale spin simulations of magnetic systems of arbitrary geometry and magnetic structure, providing a graphical user interface with powerful visualizations and an easy to use scripting interface.
An extended Heisenberg type spin-lattice Hamiltonian including competing exchange interactions between neighbors at arbitrary distance, higher-order exchange, Dzyaloshinskii-Moriya and dipole-dipole interactions is used to describe the energetics of a system of classical spins localised at atom positions.
A variety of common simulations methods are implemented including Monte Carlo
and various time evolution algorithms based on the Landau-Lifshitz-Gilbert equation of motion, which can be used to determine static ground state and metastable spin configurations, sample equilibrium and finite temperature thermodynamical properties of magnetic materials and nanostructures or calculate dynamical trajectories including spin torques induced by stochastic temperature or electric current.
Methods for finding the mechanism and rate of thermally assisted transitions include the geodesic nudged elastic band method, which can be applied when both initial and final states are specified, and the minimum mode following method when only the initial state is given.
The lifetime of magnetic states and rate of transitions can be evaluated within the harmonic approximation of transition-state theory.
The framework offers performant CPU and GPU parallelizations.
All methods are verified and applications to several systems, such as vortices, domain walls, skyrmions and bobbers are described.

\end{abstract}

\maketitle


\section{Introduction}

Multiscale materials’ simulations have emerged as one of the most powerful and widespread assets in the quest for novel materials with optimal or target properties, functionalities and performance. Simulations are employed to narrow down the design continuum of devices, to decrease the effort required to design novel materials, to substitute experiments that seem unfeasible, to suggest and analyse experiments and provide understanding of the underlying physics on scales ranging from {\AA}ngstr\"om to millimeters and from femtoseconds to decades.
In this context, spintronics is a very active field where multiscale simulations~\cite{hoffmann_antiskyrmions_2017} play an important role for the conceptualization and development of the next-generation data devices,~\cite{zutic_spintronics_2004} which includes nanoscale magnetic objects like  domain walls or  nontrivial magnetic textures such as solitons with a time dilemma of 16 orders of magnitude between writing and saving information.  Relating the requested properties to the development and choice of  magnetic materials, the simulation approach is highly useful, and a large variety of potential applications exist.~\cite{bader_colloquium_2006}
Since quantum mechanics is the key to understand magnetism, at the quantum mechanical level, \textit{ab initio} methods, such as density functional theory, can be used to calculate various properties of and interactions between atoms.
Due to the computational complexity of such calculations, they can currently only be applied to magnetic structures in crystals with length scales in the order of $1$~nm and cannot be used for time-dependent dynamics simulations on time-scales relevant for spintronics.
From \textit{ab initio} methods one may extract parameters for more approximate, atomistic spin models, such as Heisenberg type spin-lattice Hamiltonians.
There, detailed information about the electronic structure is integrated out to effective parameters describing the interaction between pairs of classical spins, so that simulations of magnetization dynamics can be extended over the timescale of nanoseconds for systems of hundreds of nanometers.
The third level of the multiscale approach in spintronics is the well-known micromagnetic approximation~\cite{brown_micromagnetics_1963} based on the assumption of a continuous magnetization vector field, defined at any point of the magnetic sample, is valid when changes of the magnetization are much larger in space than the underlying atomic lattice.
In contrast, the atomistic spin-lattice model covers the technologically increasingly important length scale from a few to several tens of nanometers. 

Here, we present a general purpose, open source, \textit{i.e.}\ publicly available, framework for atomistic spin simulations called \textit{Spirit}.~\cite{spirit}
There are actually a number of computational tools available for the simulation of the time- and space-dependent magnetization evolution.
Among the software packages for micromagnetic simulations,
two of the most impactful and widely known ones are \mbox{OOMMF}~\cite{oommf} and \mbox{mumax3}.~\cite{mumax}
This software definitely revolutionized the simulation of magnetic properties of materials and the temporal behavior of devices described by the  Landau Lifshitz Gilbert (LLG) dynamics.
Based on the micromagnetic approach these methods have well-known limitations, \textit{e.g.}\  the description of antiferromagnets, frustrated magnets, higher order non-pairwise interactions (\textit{e.g.}\ three-spin or four-spin interaction), stochastic spin dynamics and Monte Carlo simulations, \textit{etc}.
Moreover, most micromagnetic software is not interactive or provides quite limited \textit{in situ} access to the parameters of the modeled system.
Among the atomistic simulation programs, \mbox{\textit{UppASD}}~\cite{skubic_method_2008} and VAMPIRE~\cite{evans_atomistic_2014}
are examples of well-tested tools that provide important functionalities beyond LLG simulations.

The functionality of the aforementioned softwares can be greatly extended by adding an interactive graphical user interface (GUI) that can be used to control calculations in real time -- to not only change parameters, but also interact with the spin texture as demonstrated for example in Ref.~\onlinecite{rybakov_chiral_2018}.
Together with such a GUI, \textit{Spirit} unifies various computational methods that are commonly applied to atomistic (and in part also to micromagnetic) simulations: 
Monte Carlo and Landau Lifshitz Gilbert (LLG) dynamics,~\cite{nowak_thermally_2001} the geodesic nudged elastic band (GNEB) method~\cite{bessarab_method_2015},  minimum mode following~\cite{mueller_duplication_2018} (MMF), harmonic transition-state theory~\cite{bessarab_harmonic_2012} (HTST), and the visualization of eigenmodes.
All of these methods are quite distinct, but complementary in nature.~\cite{braun_topological_2012}
For example, LLG dynamics can be used 
to simulate the time evolution of a magnetic system on a short time scale, while GNEB and/or MMF can be used to find first-order saddle points of the energy landscape -- corresponding to transition states for thermally assisted transitions.
These calculations can provide important information, such as the energy barrier for the transition and can be used in HTST to calculate the lifetimes of metastable magnetic configurations over a long time scale.
The integration of these methods into a single, uniform framework 
can lead to significant increase in productivity.
The following section will introduce the structure of the \textit{Spirit} software and subsequent sections will detail the aforementioned methods, ordered by their complexity, which we rank according to the derivatives of the energy required for the implementation of these methods.
Provided examples are mainly related to magnetic skyrmions, which represents one of the most rapidly developing fields in modern nanomagnetism.
%

\section{The Framework}

\begin{figure}[!htb]
    \centering
	\includegraphics[width=0.9\linewidth]{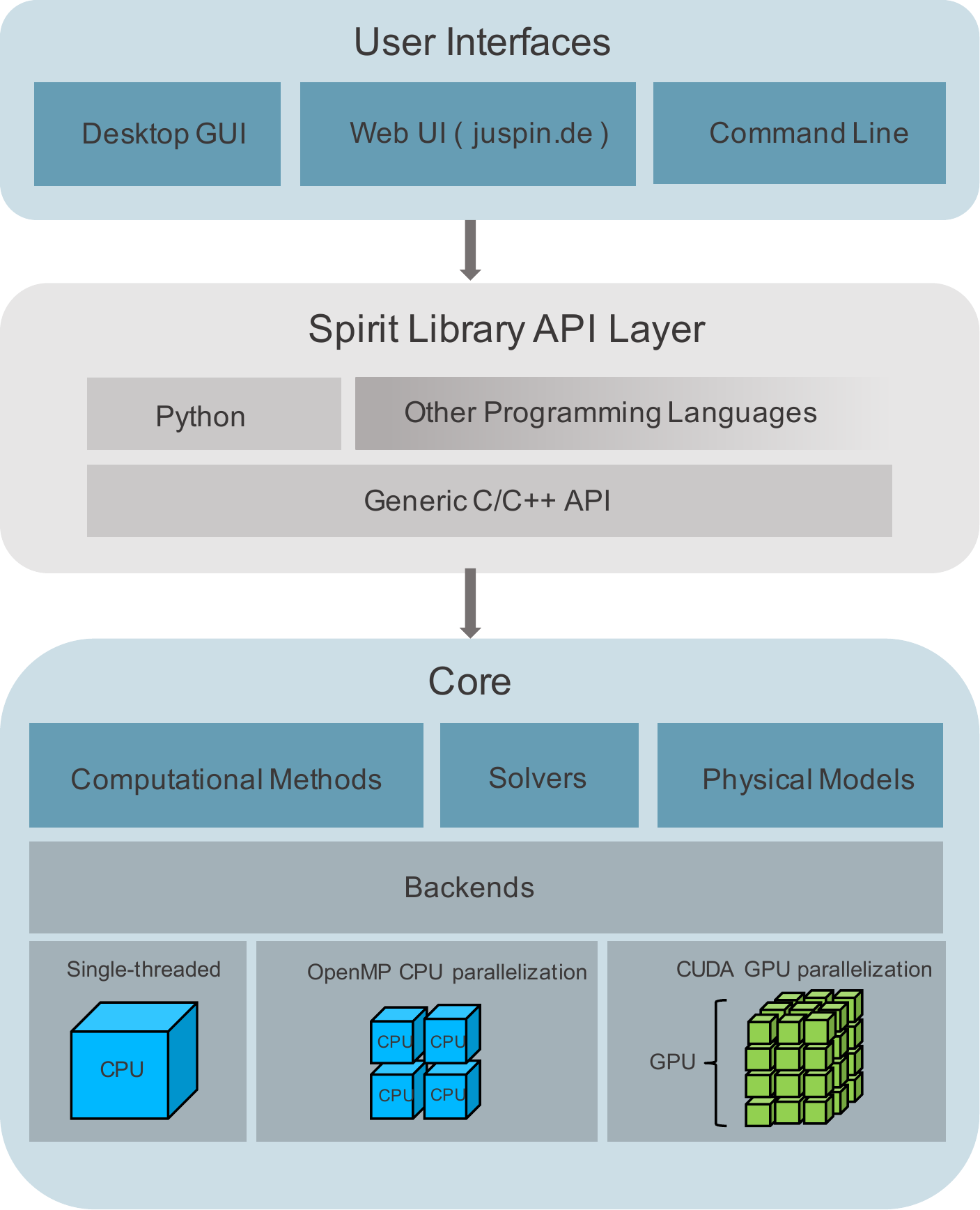}
    \caption{
    The general structure of the framework, which is separated into a core library with an application programming interface (API) layer and a set of user interfaces (UIs).
    The core library handles input/output and calculations, while the API layer provides an abstract way of interacting with the code through several programming languages.
    The UIs provide direct control of calculations, as well as real-time visualization and post-processing features.
    The back end for numerical calculations can be used in single-threaded and CPU- as well as GPU-parallel calculations.
    }
    \label{fig: framework structure}
\end{figure}

The framework consists of modular components, as illustrated in Fig.~\ref{fig: framework structure}:
a core library for calculations and input/output;
an application programming interface (API) layer to abstract the interaction with the code and provide a uniform interface across various programming languages, \textit{e.g.}\ C/C$++$ and Python;
a set of user interfaces to enable fast and easy interaction with simulations, powerful real-time visualization and post-processing features, for instance visualization of 2D and 3D magnetization vector fields with corresponding isosurfaces and visualization of eigenmodes.
The visualization of \textit{Spirit} is available as a standalone library called \textit{VFRendering},~\cite{vfrendering} which utilizes advanced features of modern OpenGL, \textit{e.g.}\ shaders, available since version 3.2.
Note that the images of spin systems in Figs.~\ref{fig: ddi test} and~\ref{fig: gneb bobber} have been generated using the graphical user interface of \textit{Spirit}.
Other examples of the visualization features of \textit{Spirit} can be found in Refs.~\onlinecite{liu_binding_2018, zheng_experimental_2018, du_interaction_2018, mueller_duplication_2018}.
\textit{Spirit} has also been used for numerical calculations in Refs.~\onlinecite{hoffmann_antiskyrmions_2017, hagemeister_controlled_2018, mueller_duplication_2018, redies_distinct_2018}.
As the API layer is written in the C programming language, many other languages can be used to call the corresponding functions and the core library can, therefore, be used in many different contexts.
An illustration of this flexibility is the implementation of an additional, web based user interface,~\cite{juspin} using JavaScript to call \textit{Spirit} and WebGL to display the simulated system.
The desktop GUI can be used to control parameters in the calculation, such as system size or interaction parameters -- useful for fast testing and setup -- as well as for direct interaction with the spin textures.
The latter is highly useful, for example to set up complex initial states~\cite{rybakov_chiral_2018} or rectify calculations, such as GNEB paths that have diverged from their intended transition
%
In order to increase productivity in repetitive or long-timescale calculations, \textit{Spirit} can be used in Python scripts.
Such scripts allow to reproduce all steps which can be taken in the GUI and therefore flexible and effective use of clusters and remote machines.
Example Python scripts can be found in the code repository~\cite{spirit}.
Note that the ability to use \textit{Spirit} from Python also enables a straightforward integration into multiscale simulations and workflow automation frameworks, such as ASE~\cite{ASE} or AiiDA.~\cite{AiiDA}
Documentation of input, features and the APIs, as well as examples of usage can be found online.~\cite{spirit}

\begin{figure}[!htb]
    \centering
	\includegraphics[width=0.9\linewidth]{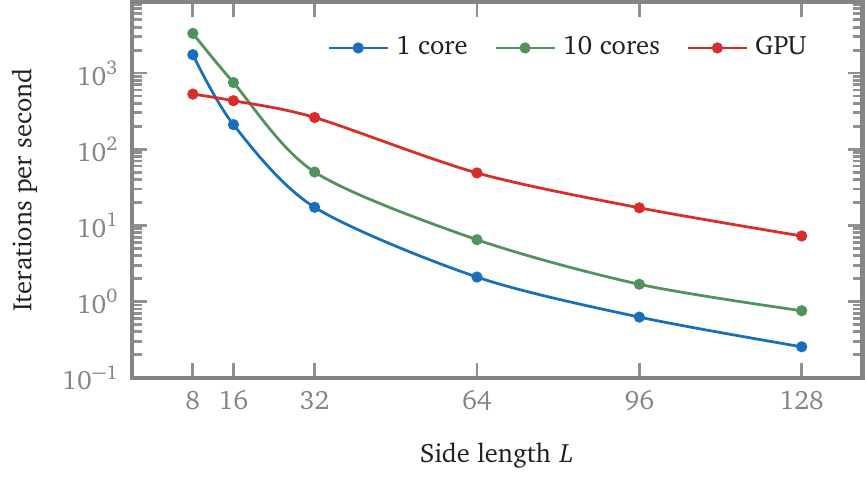}
    \caption{
    Iterations per second of a LLG simulation over side length $L$ of a simple cubic system for $1$ thread, $10$ threads and on a GPU.
    The CPU parallelization consistently increases performance by almost an order of magnitude.
    By using a GPU, another order of magnitude can be gained for large system sizes, while the GPU performance at small system sizes is limited by the overhead of CUDA kernel launches.
    Calculations were performed on a Linux system with an Intel Core i9-7900X 3.30GHz and a NVIDIA GeForce GTX 1080.
    }
    \label{fig: llg performance}
\end{figure}
%

%
The spin-lattice Hamiltonian, as well as all implemented methods and solvers have been abstracted from the specifics of numerical operations, allowing a generic \textit{backend}, which can optionally use OpenMP for CPU parallelization or CUDA for GPU parallelization.
The performance of a simple LLG simulation over different system sizes, including dipolar interactions, is shown in Fig.~\ref{fig: llg performance}.
The performance gain of the parallelization over the single-threaded case is obvious as $10$ cores give almost an order of magnitude across a broad range of system sizes and the GPU can give another order of magnitude at larger system sizes.
As expected, the speed drops with the system size.
Note that when dipolar interactions are included, due to the usage of FFTs, iterations can be slowed down if a side length of the system is not a power of two.
%

\section{Model and Methods}

\subsection{Hamiltonian}
In \textit{Spirit}, we implemented an extended Heisenberg Hamiltonian~\cite{aharoni_introduction_2000, rado_magnetism_1963} of classical spins $\vec{n}_i$ of unit length located at lattice sites $i$ giving rise to the magnetic moment $\vec{m}_i = \mu_i \vec{n}_i$.
The general form
\begin{equation}
\begin{alignedat}{1}
	\mathcal{H} =
      &- \sum_i \mu_i \vec{B}\cdot\vec{n}_i
       - \sum_i \sum_j K_j (\nvec{K}_j\cdot\vec{n}_i)^2\\
      &- \sum\limits_{\braket{ij}}\, J_{ij} \vec{n}_i\cdot\vec{n}_j
       - \sum\limits_{\braket{ij}}\, \vec{D}_{ij} \cdot (\vec{n}_i\times\vec{n}_j)\\
      &+ \frac{1}{2}\frac{\mu_0}{4\pi} \sum_{\substack{i,j \\ i \neq j}} \mu_i \mu_j \frac{(\vec{n}_i \cdot \nvec{r}_{ij}) (\vec{n}_j\cdot\nvec{r}_{ij}) - \vec{n}_i \vec{n}_j}{{r_{ij}}^3}\; ,
\end{alignedat}
	\label{eq: hamiltonian}
\end{equation}
includes
(i) the Zeeman term describing the interaction of the spins with the external magnetic field $\vec{B}$,
(ii) the single-ion magnetic anisotropy, where $\nvec{K}_j$ are the axes of the uniaxial anisotropies of the basis cell with the anisotropy strength $K_j$,
(iii) the symmetric exchange interaction given by $J_{ij}$ and the antisymmetric exchange, also called Dzyaloshinskii-Moriya interaction, given by vectors $\vec{D}_{ij}$, where $\braket{ij}$ denotes the unique pairs of interacting spins $i$ and $j$,
(iv) the dipolar interactions, where $\nvec{r}_{ij}$ denotes the unit vector of the bond connecting two spins.
Quite often, the number of pairs for the exchange interactions is limited to nearest or next-nearest neighbors only.
In Spirit the implementation of the Hamiltonian~\eqref{eq: hamiltonian} does not assume any limitation on the number of or distance between such pairs, meaning that long-ranged and non-isotropic interactions can be considered.
Additionally, higher-order multi-spin-multi-site interactions~\cite{hoffmann_systematic_2018} are implemented in \textit{Spirit} as quadruplets of the form
\begin{equation}
    E_\mathrm{Quad} = - \sum\limits_{ijkl}\, K_{ijkl} \left(\vec{n}_i\cdot\vec{n}_j\right)\left(\vec{n}_k\cdot\vec{n}_l\right)\; .
\end{equation}
These can represent the 4-spin-2-site~\cite{szilva_interatomic_2013} (also called biquadratic), the 4-spin-3-site,~\cite{kroenlein_magnetic_2018} and the 4-spin-4-site~\cite{heinze_spontaneous_2011} (also called "4-spin") interactions.
Both the system geometry and the underlying lattice symmetry can be chosen arbitrarily by setting the Bravais vectors and basis cells with any given number of atoms.
\textit{Spirit} also allows the pinning of individual spins or a set of spins, for instance belonging to the boundary layers.
One can also introduce defects, such as vacancies and atoms of different types.

\par
\textbf{Dipolar interactions.}
The dipole-dipole interaction, due to its long-ranged nature, represents the most complex contribution to the Hamiltonian~\eqref{eq: hamiltonian}.
Direct summation over all interacting spins is of complexity $\mathcal{O}(N^2)$, where $N$ is the number of spins, resulting in dramatic decrease of performance of the simulations.
By making use of fast Fourier transforms (FFTs) and the convolution theorem, the computational complexity can be reduced to $\mathcal{O}(N \log N)$.
This convolution method is well-known in micromagnetic simulations,~\cite{hayashi_ddi_1996} based on a finite difference scheme.
To treat arbitrary spin lattices with any given number of atoms in the basis cells, we use an adapted version of this method.
In particular, we consider sublattices composed of atoms with the same index in the basis cell.
One FFT is performed on each of these sublattices and additional convolutions are required to describe the interactions between the sublattices.
An efficient implementation of this scheme is achieved using high performance, robust FFT libraries.~\cite{fftw, cufft}

\begin{figure}[tb]
\centering
	\includegraphics[width=.9\linewidth]{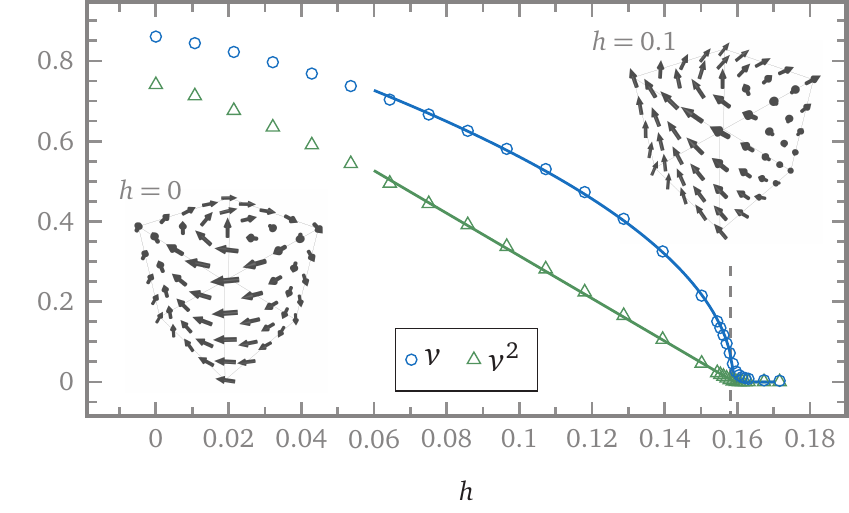}
	\caption{
	    Helicity~\eqref{eq: helicity} of a ferromagnetic cube, composed of ${50\times50\times50}$ spins on a simple cubic lattice with constant $a=1$\AA~and nearest-neighbor exchange of $J = 16.86$~meV.
        The stray-field induced helicity $\nu$ (circles) and $\nu^2$ (triangles) on the reduced external magnetic field $h$ are shown.
        $h$ is given in reduced units of $h = B/(\mu_0 M_\mathrm{s})$ with the saturation magnetisation density $M_\mathrm{s}$.
        The fitted curves (solid lines) show that the dependence of $\nu^2$ close to the critical field is approximately linear and they give a critical field value of $h_\mathrm{c}=0.159$ -- which matches the expected value of $h_\mathrm{c}=0.158$, as shown in Ref.~\onlinecite{hubert_systematic_1999}, within $1\%$.
        The two insets show illustrations of how the cube will be magnetized at $h=0$ (left) and $h=0.1$ (right).
    }
\label{fig: ddi test}
\end{figure}

To verify our implementation of dipolar interactions, we compared it to the direct evaluation of the sum for random configurations with spatially non-symmetric basis cells and checked the convergence of the stray field of a homogeneously magnetized monolayer against the analytically known result.
Here, we show that \textit{Spirit} correctly reproduces the solution of typical problems, \textit{e.g.}\ Ref.~\onlinecite{hubert_systematic_1999}, by calculating the stray field-induced helicity of a ferromagnetic cube.
The helicity is defined as the absolute value of the line integral over the curve $C$ which is composed of the upper edges of the cube:
\begin{equation}
	\nu =  \frac{\left|\oint_{C} \vec{n} \cdot \mathrm{d}\vec{s}\right|}{\oint_{C} |\mathrm{d}\vec{s}|} \; .
\label{eq: helicity}
\end{equation}
In the atomistic case this is discretized into the appropriate sums.
The energy minimization was performed using a Verlet-like velocity projection method (see Appendix~\ref{app: vp}).
The results are shown in Fig.~\ref{fig: ddi test}.
The squared helicity is expected to approach the critical field linearly, so that a line can be fitted to extract the precise result from the calculations.
We note that the resulting critical field converges to the expected value of $h_\mathrm{c}=0.158$ with increasing resolution of the grid, where a cube with $30^3$ lattice sites already gives an agreement within $2\%$ and the shown example with $50^3$ sites a discretization error of less than $1\%$ with respect to the continuum solution.

\par
\textbf{Topological charge.}
In the past years, we witnessed the characterization of smooth magnetization fields $\vec{m}(\vec{r}) = \mu \vec{n}(\vec{r})$ in terms of topological classes.
In this case, the $\mathbb{S}^2$ winding number defined as
\begin{equation}
    Q(\vec{n})= \frac{1}{4\pi} \int_{\mathbb{R}^2} \vec{n} \cdot (\partial_x \vec{n} \times \partial_y \vec{n})\, \mathrm{d}\vec{r} \; .
\label{eq: topocharge}
\end{equation}
It should be noted that a secondary topological charge $v$ can be considered as 
the defining index to distinguish between skyrmion ($v=1$) and antiskyrmion ($v=-1$) independently of the background state.
It is defined as
\begin{equation}
	v = \frac{1}{2\pi} \oint_{\Gamma} \frac{n_x \mathrm{d}n_y - n_y \mathrm{d}n_x}{n_x^2 + n_y^2}
\label{eq: topovorticity}
\end{equation}
and is also referred to as the $\mathbb{S}^1$ winding number in comparison to the conventional topological charge $Q$.
Here, $\Gamma$ denotes an oriented Jordan curve around the center of the skyrmion.

The notion of skyrmion/antiskyrmion refers to a local energy minimizer of \eqref{eq: hamiltonian} within the topological class characterized by the relative skyrmion number $N=\pm 1$.
For sufficiently regular fields $\vec{n}$ decaying to the background state $n_z(\infty) =\pm 1$, the index $N$ is defined relative to this background state, $N(\vec{n})=-n_z(\infty) Q(\vec{n})$.\cite{hoffmann_antiskyrmions_2017}
In a typical situation where the horizontal magnetization field vanishes at a single point (skyrmion center) the relative skyrmion number $N$ agrees with the index of the horizontal magnetization field at the skyrmion center.
It is customary to fix the background state $n_z(\infty)=1$, which leads to the characterization $Q=-1$ for skyrmions and $Q=+1$ for antiskyrmions.

The evaluation of expression \eqref{eq: topocharge} for the spindensity $\vec{n}$ on a discrete lattice, as implemented in \textit{Spirit}, is outlined in Appendix~\ref{app: topocharge}.
%

%
\subsection{Monte Carlo}
The Monte Carlo method is well-known in Physics and has a broad range of applications.~\cite{binder_monte_1981}
We have implemented a basic Metropolis algorithm with a cone angle for the displacement of the spins.~\cite{hinzke_monte_1999, nowak_thermally_2001}
This requires only the calculation of the energy, making it the most straightforward of the methods implemented in \textit{Spirit}.
While it is a useful tool to calculate equilibrium properties, the drawback is that it cannot resolve time-dependent processes.
One iteration of the Metropolis algorithm will sequentially -- but in random order -- pick each spin in the system once and perform a trial step.
Trial steps are preformed by defining a relative basis in which the current spin is the $z$-axis and choosing a new spin direction by uniformly distributed random variables $\varphi\in[0,2\pi]$ and $\cos(\theta)\in[0,\cos(\theta_\mathrm{cone})]$, where $\theta_\mathrm{cone}$ is the opening angle of the cone.
The trial step is accepted with a probability
\begin{equation}
    P = e^{-\Delta E/k_\mathrm{B}T}\; ,
\end{equation}
where $\Delta E$ is the energy difference between the previous spin configuration and the trial step.
The cone angle can be set by an adaptive feedback algorithm according to a desired acception-rejection ratio.
\begin{figure}
\centering
	\includegraphics[width=.9\linewidth]{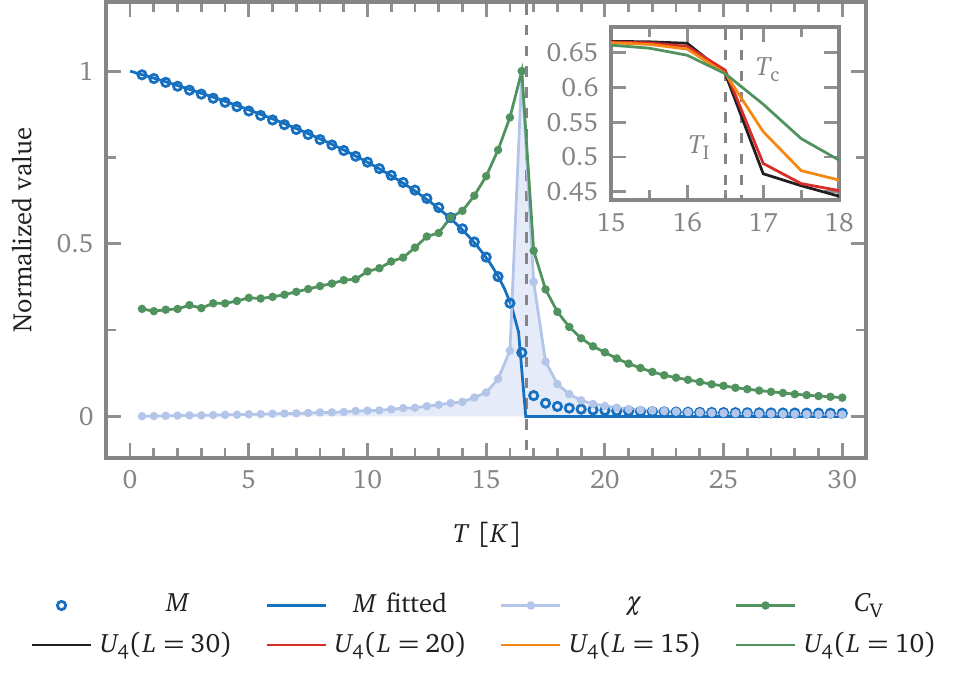}
	\caption{
	    A ${30\times 30\times 30}$ ferromagnet with $J=1$~meV, with an expected critical temperature of $T_\mathrm{C} \approx 16.71$~K.
        Normalized values of
        the total magnetization $M$, susceptibility $\chi$, specific heat $C_\mathrm{V}$ and $4^\mathrm{th}$ order Binder cumulant $U_4$ are shown.
        The magnetization is fitted with $M(T) = \left( 1 - T/T_\mathrm{c} \right)^b$.
		At each temperature, $10$k thermalisation steps were made before taking $100$k samples.
        Monte Carlo calculations give $T_\mathrm{c} \approx 16.60$~K -- an agreement with expectation within $1\%$. The exponent is fitted with $b\approx0.33$.
		The inset shows the Binder cumulants for system sizes of $L=30$, $L=20$, $L=15$ and $L=10$, giving an intersection at $T_\mathrm{I}=16.5+-0.25$, which is an excellent agreement with the expected value of $T_\mathrm{c}$ within the temperature step of $0.5$~K.
    }
	\label{fig: monte carlo}
\end{figure}
Using this method, one can, for example, calculate the critical temperature of a spin system.
It is known that the isothermal susceptibility is related to the magnetization fluctuations~\cite{landau_guide_2009}
\begin{equation}
	\chi = \frac{1}{k_\mathrm{B}T}\left( \braket{m^2} - \braket{m}^2 \right)\; ,
\end{equation}
where $m = \frac{1}{N} |\sum_i \vec{n}_i|$ is the average magnetization of the sample, while the specific heat relates to fluctuations of the energy
\begin{equation}
	C_\mathrm{V} = \frac{1}{k_\mathrm{B}T^2}\left( \braket{E^2} - \braket{E}^2 \right)\; ,
\end{equation}
where both should diverge at the critical temperature $T_\mathrm{c}$ for a phase transition, \textit{e.g.}\ to the paramagnetic phase.
The $4^\mathrm{th}$ order Binder cumulant~\cite{binder_finite_1981}, which is often used to avoid finite size scaling effects, is defined as
\begin{equation}
	U_4 = 1 - \frac{\braket{m^4}}{3\braket{m^2}^2} \; .
\end{equation}
Fig.~\ref{fig: monte carlo} shows these quantities as results of a Monte Carlo calculation of a cube of ${30\times30\times30}$ lattice sites for $J=1$~meV.
For a simple cubic ferromagnet, from the high-temperature expansion method,~\cite{baker_high-temperature_1967} the critical temperature is known to be~\cite{rocio_modeling_2011} $T_\mathrm{c} = 1.44~J/k_\mathrm{B} = 16.71$~K.
The results shown in Fig.~\ref{fig: monte carlo} demonstrate the validity of the implementation, as the expected critical temperature is matched with an error of less than $1\%$.
Note that in Monte Carlo methods, the parallel tempering algorithm has proven to be an effective tool.~
\cite{swendsen_replica_1986, hukushima_exchange_1996, bottcher_b-t_2017}
The usage of Python and a MPI package would enable one to quite easily reproduce this algorithm in a Python script using \textit{Spirit}.
%
%
%
\subsection{Landau-Lifshitz-Gilbert Dynamics}
%
%
%
The Landau-Lifshitz-Gilbert (LLG) equation~\cite{landau_on_1935, gilbert_phenomenological_2004} is the well-established equation of motion for the dynamical propagation of classical spins.
In its explicit form and including spin torque and temperature contributions,~\cite{brown_thermal_1963, schieback_numerical_2007} it can be written
\begin{equation}
    \begin{alignedat}{2}
    \dfrac{\partial \vec{n}_i}{\partial t}
        =& - \dfrac{\gamma}{(1+\alpha^2)\mu_i} \vec{n}_i \times \vec{B}^\mathrm{eff}_i \\
        &- \dfrac{\gamma \alpha}{(1+\alpha^2)\mu_i} \vec{n}_i \times (\vec{n}_i \times \vec{B}^\mathrm{eff}_i) \\
        &- \dfrac{\alpha-\beta}{(1+\alpha^2)} u \vec{n}_i \times (\vec{j}_e \cdot \nabla_{\vec{r}} )\vec{n}_i \\
        &+ \dfrac{1+\beta \alpha}{(1+\alpha^2)} u \vec{n}_i \times (\vec{n}_i \times (\vec{j}_e \cdot \nabla_{\vec{r}} )\vec{n}_i) \; ,
    \end{alignedat}
\label{eq: llg}
\end{equation}
in which the terms correspond to (i) precession, (ii) damping, (iii) precession-like current-induced spin torque, and (iv) damping-like current-induced spin torque.
$\gamma$ is the electron gyromagnetic ratio, $\alpha$ is the damping parameter, $\vec{B}^\mathrm{eff}_i$ is the effective field, $\beta$ is a non-adiabatic parameter, $\vec{j}_e$ is the electron current vector, and $\nabla_{\vec{r}} = \partial / \partial \vec{r}$ is the spatial gradient acting on the spin orientation.
The effective field always contains a component related to the energy gradient $\vec{B}^\mathrm{eff}_i = - \partial \mathcal{H} / \partial \vec{n}_i$, but in this notation for the LLG equation, the effective field may contain also a stochastic thermal field, \textit{i.e.}\ $\vec{B}_i \to \vec{B}_i + \vec{B}_i^\mathrm{th}$, given by
\begin{equation}
    \vec{B}^\mathrm{th}_i(t,T) = \sqrt{2D_i(T)} \vec{\eta}_i(t) = \sqrt{2\alpha k_\mathrm{B}T \frac{\mu_i}{\gamma}} \vec{\eta}_i(t) \; ,
\end{equation}
where the magnitude is given by the fluctuation-dissipation theorem and $\vec{\eta}_i$ is white noise, such that $\braket{\vec{B}_{i\alpha}^\mathrm{th}(t,T)}_t = 0$ and $\braket{\vec{B}_{i\alpha}^\mathrm{th}(t,T) \vec{B}_{j\beta}^\mathrm{th}(0,T)}_t = 2D_i(T) \delta_{ij}\delta_{\alpha\beta}\delta(t)$.
To achieve these properties in an implementation, the vectors $\vec{\eta}_i(t)$ can each be created from three independent standard normally distributed random values at every time step.
Note also that in time-integration schemes, to fulfill the fluctuation-dissipation relation, the thermal field needs to be normalized by the time step with a factor $1/\sqrt{\delta t}$.
For more details on the integration of the stochastic LLG equation see for example references~\onlinecite{mentink_stable_2010,bauer_thermally_2011,levente_langevin_2014} and references therein.

Sampling of the stochastic LLG for the same parameters as shown in Fig.~\ref{fig: monte carlo} is presented in Appendix~\ref{app: langevin}, verifying the implementation and the equivalence of the stochastic LLG and Monte Carlo methods.

\begin{figure}[t]
	\begin{minipage}{\linewidth}
	\includegraphics[width=.9\textwidth]{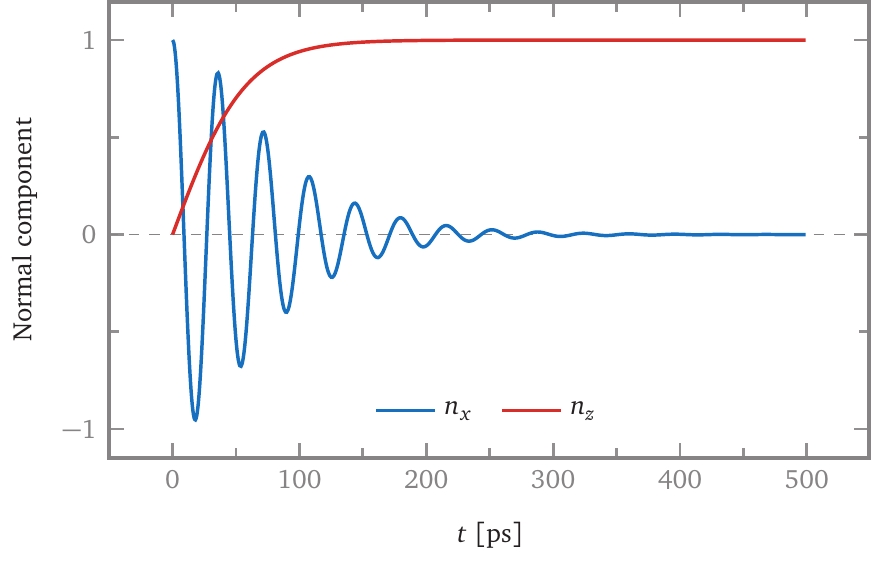}
	\end{minipage}%

	\begin{minipage}{\linewidth}
	\includegraphics[width=.9\textwidth]{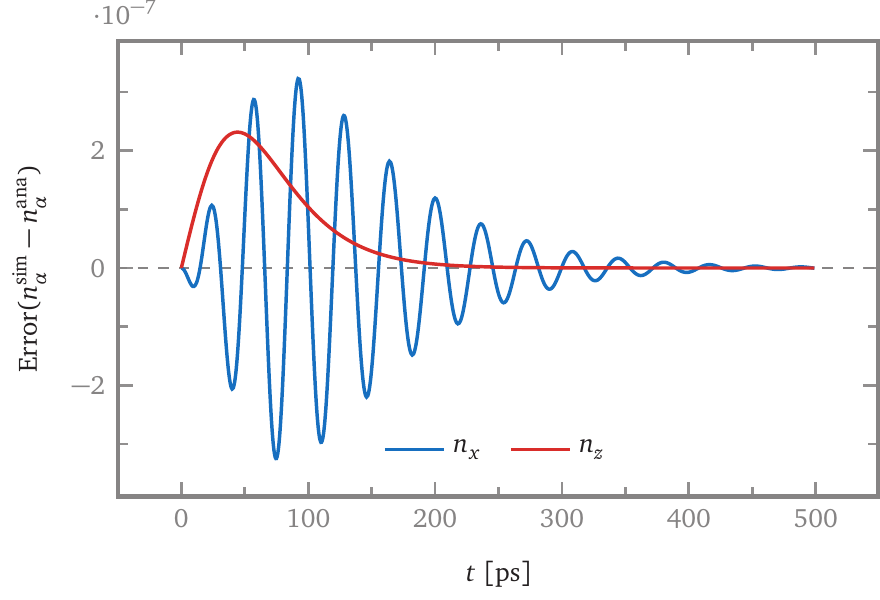}
	\end{minipage}%
    
	\caption
    {
    LLG calculation of a single spin in an external magnetic field of $B=1$~T with a damping of $\alpha=0.1$ and a timestep of $\mathrm{d}t=10$~fs, using the Depondt method.
    Note that the error may depend strongly on the time step and damping.
    While the Heun method matches well with results shown in Ref.~\onlinecite{evans_atomistic_2014}, giving an error within $10^{-6}$, the Depondt method shows a lower error of around $3\times10^{-7}$ with respect to the analytical solution.
    }
	\label{fig: llg precession}
\end{figure}

\begin{figure}
	\begin{minipage}{\linewidth}
	\centering
	\hspace{0.1\linewidth}\includegraphics[width=.7\linewidth]{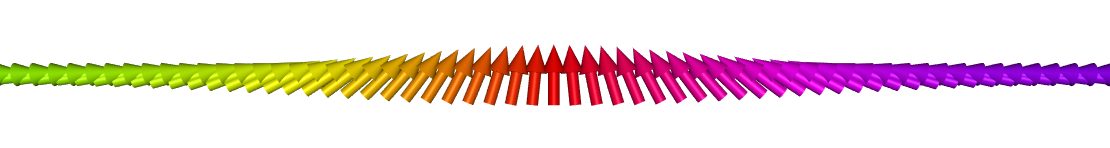}
	\end{minipage}%

	\begin{minipage}{\linewidth}
	\includegraphics[width=.9\linewidth]{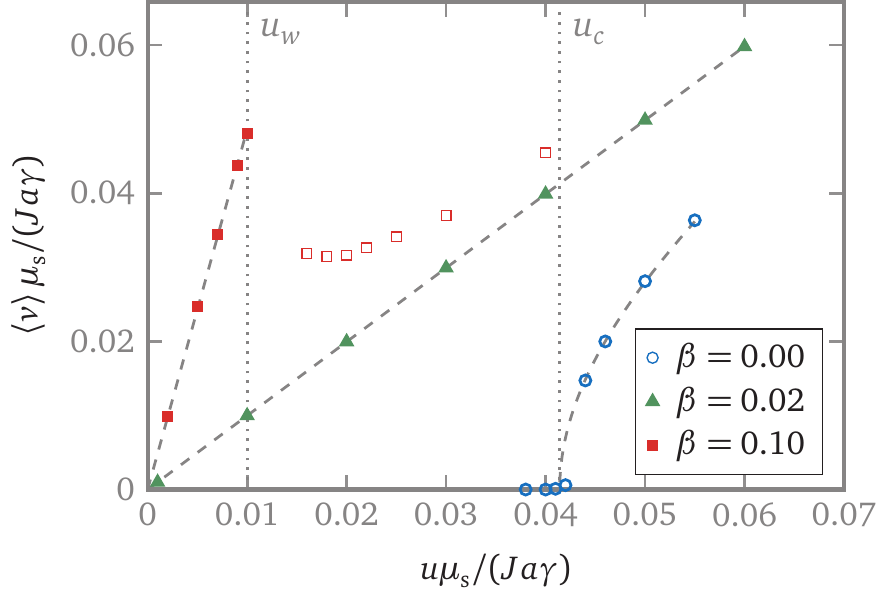}
	\end{minipage}%

	\caption{
	The average velocity of head-to-head domain wall (see top) for various values of the non-adiabatic parameter $\beta$.
	For $\beta = 0.10$ the Walker breakdown occurs at approximately $u_W \approx 0.01$.
	For $\beta = 0$ a critical current is at $u_c \approx 0.0414$. From this point the relation $\langle v \rangle = \sqrt{u^2-u_c^2}/(1+\alpha^2)$ mentioned by Thiaville {\itshape et al.}~\cite{thiaville_domain_2004} takes effect.
	The mentioned relation is fitted to the data for $\beta = 0$.
	For $\beta = 0.1$ and currents under the walker breakdown and $\beta = 0.02$ the dashed lines show linear fits.
	Open symbols denote rotation around the $x$-axis.
	The results from Ref.~\onlinecite{schieback_numerical_2007} are reproduced well.
	}
	\label{fig:dwvelocity}
\end{figure}

In order to evolve a spin system in time according to this equation, quite a few well-known solvers can be applied.
In \textit{Spirit}, currently Heun's method,~\cite{nowak_thermally_2001} a $4^\mathrm{th}$ order Runge-Kutta solver, Depondt's Heun-like method~\cite{depondt_spin_2009}, and Mentink's semi-implicit method B (SIB)~\cite{mentink_stable_2010} are implemented (see Appendices \ref{app: heun depondt} and \ref{app: sib} for details).
These methods can also be used for energy minimization by considering only the damping part of the LLG equation.
However, experience has shown that a Verlet-like velocity projection solver~\cite{bessarab_method_2015} can greatly improve convergence to the closest energy minimum, as it carries a fictive momentum (see Appendix~\ref{app: vp} for details).
An easy test for the validity of the implemented dynamics solvers is the Larmor precession and the damping of a single spin in an external magnetic field, as shown in Fig.~\ref{fig: llg precession}.
The analytical equations with which the results can be compared are
\begin{equation}
    \begin{alignedat}{1}
        n_z(t)     &= \tanh\left( \frac{\alpha\gamma}{(1+\alpha^2)\mu} |\vec{B}| t \right) \\
        \varphi(t) &= \frac{\gamma}{(1+\alpha^2)\mu} |\vec{B}| t \\
        n_x(t)     &= \cos(\varphi(t)) \sqrt{1-n_z^2(t)}\; .
    \end{alignedat}
\end{equation}
The errors of the Depondt solver, shown in Fig.~\ref{fig: llg precession}, match those of an equivalent calculation given in Ref.~\onlinecite{evans_atomistic_2014}.
In order to verify our implementation of spin current induced torques, the results from Ref.~\onlinecite{schieback_numerical_2007} on the velocity of a domain wall in a head-to-head spin chain were reproduced for various non-adiabatic parameters $\beta$.
The chain is oriented along the $x$-axis and the first and the last spin are fixed in $+x$ and $-x$ direction, respectively.
As a subset of the general Hamiltonian~\eqref{eq: hamiltonian}, the Hamiltonian for this example can be written as follows:
\begin{equation}
    \mathcal{H} =
        -\sum_i K_1 n_{ix}^2 + K_2 n_{iy}^2
        -J \sum_{\langle ij \rangle} \vec{n}_i \cdot \vec{S}_j\; .
\label{eq: dw hamiltonian}
\end{equation}
The reference provides analytical equations against which the numerical results were checked.
In Fig.~\ref{fig:dwvelocity} we show the data for the average domain wall velocity $\braket{v}$ over applied current $u$ in normalised units.
The approximate prediction~\cite{thiaville_domain_2004} $\langle v \rangle = \sqrt{u^2-u_c^2}/(1+\alpha^2)$ fits the results well, as shown in Fig.~\ref{fig:dwvelocity}.
As expected, we observe the Walker breakdown~\cite{schryer_motion_1974} and a critical effective velocity of $u_c \approx 0.0414$, which is in close agreement with the reported value of $u_c \approx 0.0416$.
Note, for $\beta = 0.1$ and currents larger than $u_W$ and for $\beta = 0$ and currents larger than $u_c$, the wall starts rotating around the $x$-axis.
%
%
%
\subsection{Geodesic Nudged Elastic Band Method}
When determining the rates of some rare transition events or the lifetimes of metastable magnetic states, LLG dynamics simulations typically are typically unfeasible due to the disparity between the time scales of the simulation and the transition events.
An approach to this problem is given by a set of rate theory methods, namely the geodesic nudged elastic band~\cite{bessarab_method_2015} (GNEB) and minimum mode following~\cite{mueller_duplication_2018} (MMF) methods together with harmonic transition-state theory~\cite{bessarab_harmonic_2012} (HTST).
The latter two are higher order methods, requiring knowledge of the second derivatives of the energy -- the Hessian matrix -- and will be described in the following sections.
The GNEB method is a way of calculating minimum energy transition paths between two pre-determined configurations.
The path is discretized by a number of spin configurations, in the following called images.
In order to converge from an initial guess to a stable, energy-minimized path, spring forces are applied along the path tangents, while energy gradient forces are applied orthogonal to the path tangents.
The total force therefore reads
\begin{equation}
    F^\mathrm{tot}_\nu = F^\mathrm{S}_\nu + F^\mathrm{E}_\nu\; ,
\end{equation}
where $\nu$ is the image index along the chain, $F^\mathrm{S}$ is a spring force, and $F^\mathrm{E}_\nu$ is an energy gradient force.
The forces in this section are $3N$-dimensional vectors.
A simple definition of the spring force, which gives an equidistant distribution of images in phase space, is given by
\begin{equation}
	F^\mathrm{S}_\nu = (l_{\nu-1,\nu}-l_{\nu,\nu+1})\ \tau_\nu\; ,
\label{Eq:Spring_Force}
\end{equation}
where $l_{\nu,\mu}$ is a measure of distance between images $\nu$ and $\mu$ and $\tau_\nu$ is the (normalized) path tangent at image $\nu$.
The $F_\nu^\mathrm{E}$ should pull each image towards the minimum energy path, while leaving the distance to other images unchanged.
They can be defined to be orthogonal to the path by orthogonalizing with respect to the tangents
\begin{equation}
	F^\mathrm{E}_\nu = -\nabla E_\nu + (\nabla E_\nu \cdot \tau_\nu)\tau_\nu\; ,
\end{equation}
where $\nabla_i = \partial/\partial\vec{n}_i$.
The path tangents can be easily approximated by finite differences between spin configurations, but in order to avoid the formation of kinks in the path the definitions given in Ref.~\onlinecite{henkelman_improved_2000} should be used.

\begin{figure}[bt]
	\begin{minipage}{\linewidth}
    \centering
    \hspace{0.1\linewidth}\includegraphics[width=0.7\linewidth]{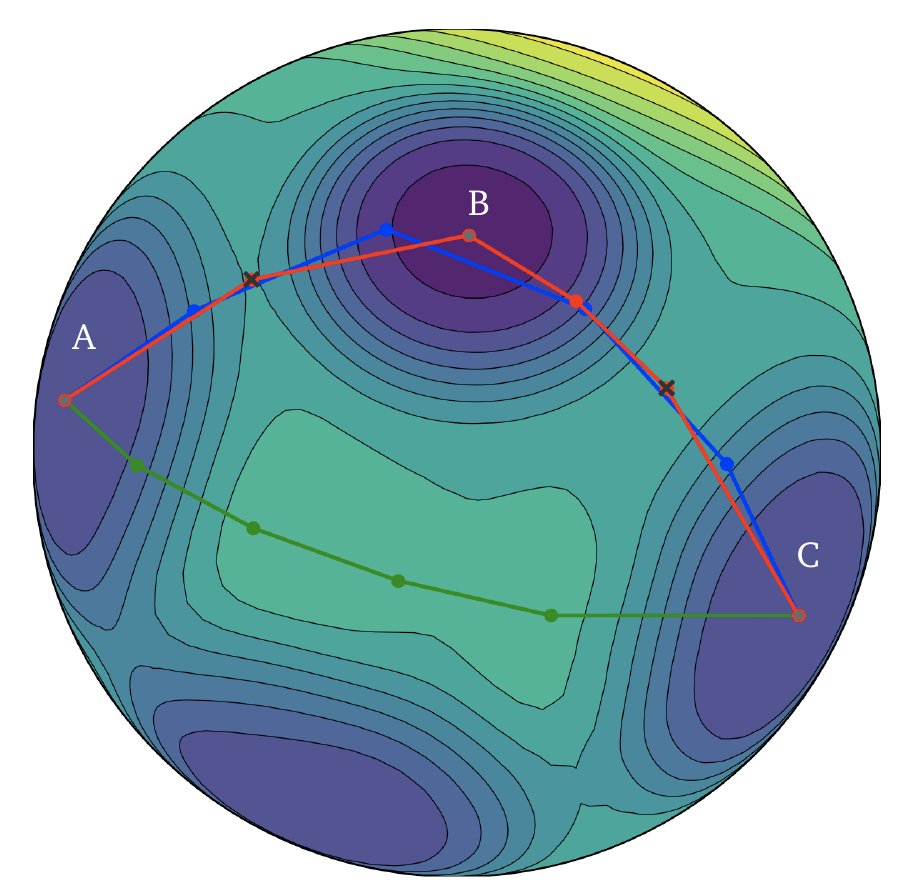}
	\end{minipage}%

	\begin{minipage}{\linewidth}
    \includegraphics[width=0.9\linewidth]{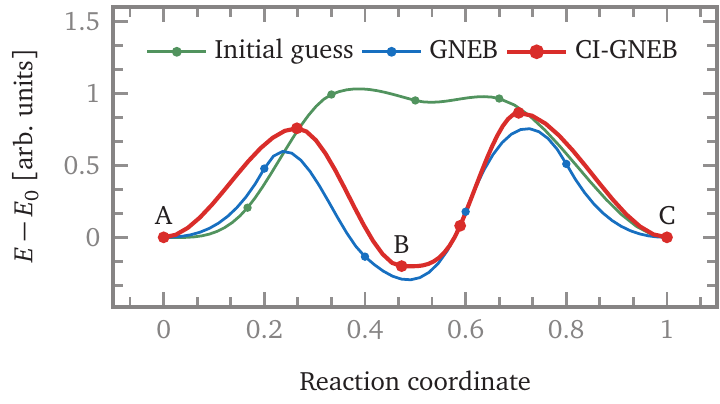}
	\end{minipage}%

    \caption{
    An illustration of the GNEB method for a single spin system (the Hamiltonian and corresponding parameters are given in Appendix~\ref{app: gneb gaussian}).
    The two-dimensional energy landscape is shown superimposed on a unit sphere.
    The initial guess (green), relaxed path (blue), and final path using climbing and falling images (red) are shown.
    }
    \label{fig: gneb single spin}
\end{figure}
In order to precisely find the point of highest energy along the minimum energy path, a first order saddle point of the energy landscape, one can use a so-called climbing image.~\cite{henkelman_climbing_2000}
Convergence onto the saddle point is achieved through the deactivation of the spring force for that image, while inverting the energy gradient force along the path:
\begin{equation}
	F^\mathrm{S,CI}_\nu = 0,
	\quad \quad
	F^\mathrm{E,CI}_\nu = - \nabla E_\nu + 2 (\nabla E_\nu \cdot \tau_\nu) \tau_\nu\; .
\end{equation}
This will cause it to minimize all degrees of freedom, except the tangent to the path, which is instead maximized.
So far, the definitions match those of the regular NEB method. In order to use the NEB method for spin systems, it is necessary to consider the constraint of constant spin length and treat tangents and force vectors accordingly.~\cite{bessarab_method_2015}
For more details see Appendix~\ref{app: gneb tangents} and \ref{app: manifold}.
In order to verify and illustrate the GNEB method, we show the example of a single spin in a set of Gaussian potentials (see Appendix~\ref{app: gneb gaussian}).
Fig.~\ref{fig: gneb single spin} shows the initial guess, made by homogeneous interpolation between the initial and final configuration, as well as a relaxed chain of images and a chain with two climbing and one falling image.
The climbing images converge onto the saddle points and the falling image onto an additional local minimum, so that the energy barriers are known exactly.
\begin{figure}[tb]
	\includegraphics[width=0.9\linewidth]{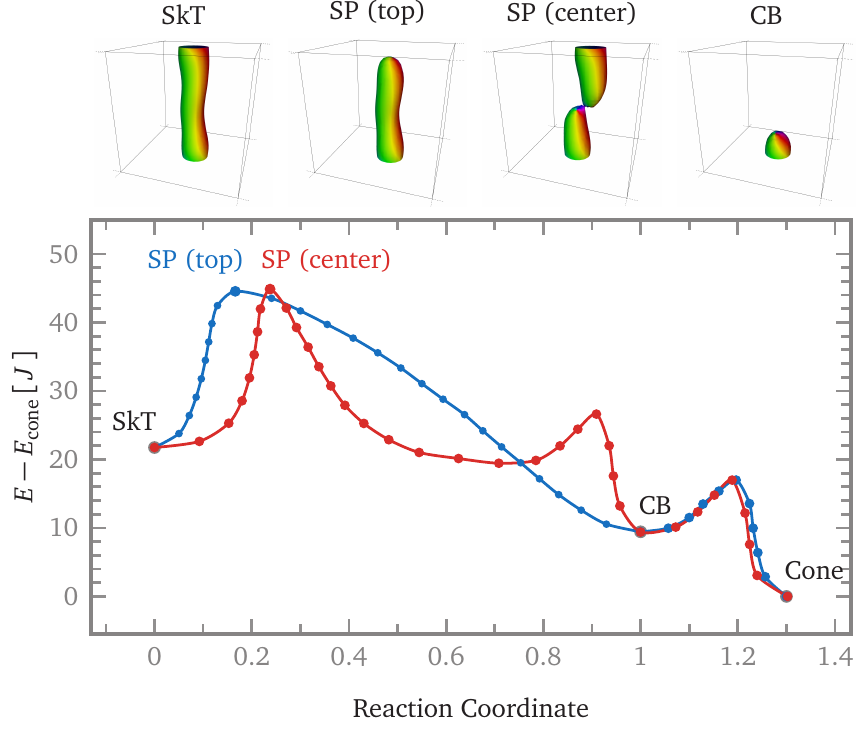}
	\caption
	{
      The skyrmion tube (SkT) is either cut in half by the nucleation of a pair of Bloch points in the center (red MEP) or separated from the upper surface by nucleation of a single Bloch point (blue MEP).
      At a field strength of $H=0.8~H_\mathrm{D}$, both processes have almost equal energy barriers of $\Delta E_\mathrm{center}=23.13~J$ and $\Delta E_\mathrm{surface}=22.81~J$.
      A chiral bobber is formed (two when the skyrmion tube is cut in half), whose collapse has an energy barrier of $\Delta E_\mathrm{bobber}=7.55~J$.
      Note that the slight differences in the collapse of the CB between the two paths come from different initial paths.
    }
	\label{fig: gneb bobber}
\end{figure}
The implementation of the GNEB method can be further tested using a conceptually simple process, which has enough degrees of freedom to pose a challenge for convergence:
the destruction of a skyrmion tube in a chiral magnetic thin film.
The parameter set is chosen in accordance with a calculation presented in Ref.~\onlinecite{rybakov_new_2015}, where a novel particle-like state is shown to emerge along the minimum energy path -- the chiral bobber.
The nucleation of a pair of Bloch points, cutting the skyrmion tube in half, is reported, resulting in the formation of one chiral bobber at each surface of the film.
In fact, as we show in Fig.~\ref{fig: gneb bobber}, also a single Bloch point can be nucleated at one of the films free surfaces.
For these calculations the specific parameters are $J=1$ and $D=0.45~J$, meaning that the incommensurate spin spiral has a period of $L_\mathrm{D} = 13.96~a$.
We note that the conical phase background -- corresponding to the ground state of the system -- introduces additional modes with little energy cost associated and this can slow the rate of convergence to the minimum energy path.
The climbing-image method~\cite{henkelman_climbing_2000} was used to converge nearby images onto the maxima along the path and -- analogous to what is suggested in the reference -- the spring force was modulated to distribute images evenly along the energy curve.
The latter improves the convergence onto the maxima, as the resolution for the finite-difference calculation of the tangents at the saddle points is increased.
As it is common to calculate cubic polynomials to interpolate between the discrete points, the segment length of these polynomials can be used for the spring forces between the images.
In \textit{Spirit}, an additional parameter is implemented, with which one can set the weighting of energy versus reaction coordinate.
Without the climbing image method, energy barrier calculations may be quite imprecise, especially when the resolution near the maximum is low. 
This is illustrated by the fact that we observe a ratio of the energy barriers between the collapse of the bobber and the Bloch point nucleation of only $3.3$, while Ref.~\onlinecite{rybakov_new_2015} -- not using climbing images -- reports a ratio of $4.3$.
\begin{figure}[tb]
	\includegraphics[width=0.9\linewidth]{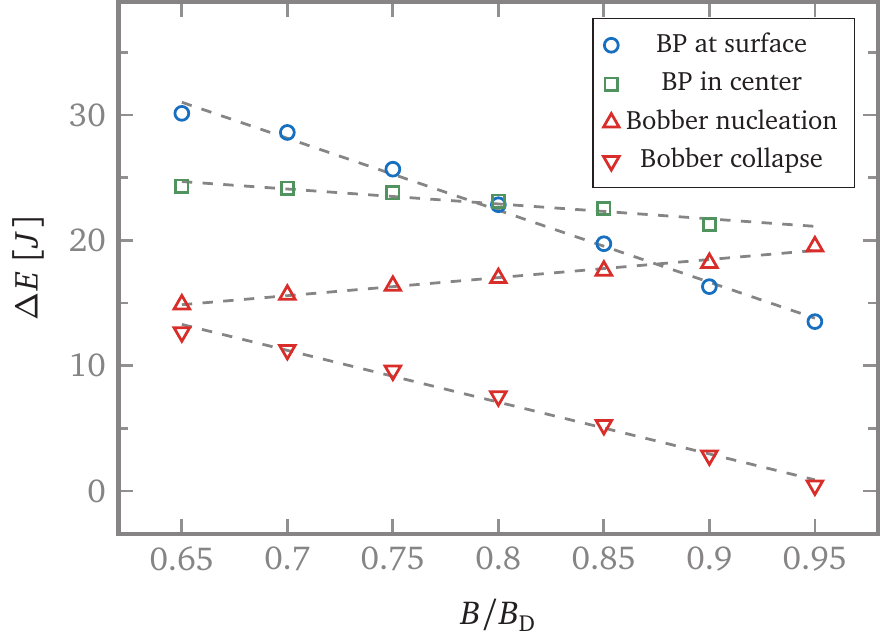}
	\caption
	{
        Energy barriers for the nucleation of Bloch points at the surface (blue circles) and in the center (green square), as well as the nucleation (red triangles up) and collapse (red triangles down) of a chiral bobber for a cube of size ${30\times30\times30}$ over applied magnetic field $H$.
		Periodic boundary conditions are applied in the $xy$-plane.
		The BP nucleation at the surface and center represents collapse of a skyrmion tube, while the bobber nucleation represents the creation of a BP in an otherwise homogeneous sample.
    }
	\label{fig: bobber barriers}
\end{figure}
The GNEB calculations reveal a crossover between the two Bloch point nucleation mechanisms, where at increasing field it becomes favorable to nucleate just one Bloch point at the surface.
It can further be seen that the energy barrier for the collapse of the bobber goes to zero right below the critical field $H_\mathrm{D}$, meaning that -- in the frame of this model -- it can only be stabilised in the conical phase.
In order to give additional quantitative reference results for this parameter set, the dependence of the energy barrier on the external magnetic field is also presented
in Fig.~\ref{fig: bobber barriers}.
%
%
%
%
\subsection{Harmonic Transition-State Theory}
As certain processes may be too rare or the desired time scale, which is to be simulated, too large to allow for dynamical simulations, other approaches are essential in estimating stability and the calculation of lifetimes of metastable magnetic states.
One can employ the well-known transition-state theory,~\cite{wigner_transition_1938} which has been used extensively, \textit{e.g.}\ in chemical reaction and diffusion calculations.~\cite{truhlar_current_1996}
The rate of transitions can be estimated from the probability of finding the system in the most restrictive and least likely region separating the initial state from possible final states -- the transition state, sometimes also called dividing surface.
Within the harmonic approximation to transition-state theory~\cite{bessarab_harmonic_2012} (HTST), one can make simplifications allowing the analytical calculation of the rate, which is then given by an Arrhenius-type law with an exponential dependence on the inverse temperature $T$ and the energy barrier of the transition $\Delta E$:
\begin{equation}
    \Gamma^\mathrm{HTST} = \frac{v}{2\pi} \Omega_0 e^{-\Delta E/k_\mathrm{B}T}\; ,
\end{equation}
where
\begin{align}
    \Omega_0
        &= \sqrt{\frac{\det^\prime H^\mathrm{M}}{\det^\prime H^\mathrm{S}}}
        = \sqrt{\frac{\sideset{}{'}\prod_i \lambda_i^\mathrm{M}}{\sideset{}{'}\prod_i \lambda_i^\mathrm{S}}}\; , \\
    \label{eq: entropy prefactor}
    v
        &= \sqrt{ 2\pi k_\mathrm{B}T }^{N_0^\mathrm{M} - N_0^\mathrm{S}}
		\frac{V^\mathrm{S}}{V^\mathrm{M}}
		\sqrt{\sideset{}{'}\sum_i \frac{a_i^2}{\lambda_i^\mathrm{S}}}\; ,
\end{align}
where the M and S superscripts indicate the minimum and first order saddle point of the transition.
The $\lambda_i$ are eigenvalues of the Hessian matrix (see Appendix~\ref{app: manifold}), $V$ are the phase space volumes of zero modes (if present, otherwise $V=1$), $N_0$ are the number of zero modes -- modes with zero eigenvalue -- and $a_i$ are coefficients in the expansion of the velocity along the unstable mode.
The primes next to determinants, products, and sums denote that only positive eigenvalues are taken into account.
The factors $a_i$ are in fact velocities: the first row of the dynamical matrix $\mathcal{V}$ transformed into the eigenbasis of the Hessian according to
\begin{equation}
	\mathcal{V} |^{2N} = \Lambda^T T^T \mathcal{V} |^{3N} T \Lambda \;,
\label{eq: dynamical matrix 2N}
\end{equation}
where $T$ is a ${3N\times2N}$ basis matrix of the tangent space and $\Lambda_{ij}$ denote the matrix of the Hessians eigenvectors (in $2N$-representation, \textit{i.e.}\ in the basis $T$).
See Appendix~\ref{app: manifold} for more information.
%

%
\begin{figure}[tb]
    \centering
	\includegraphics[width=0.9\linewidth]{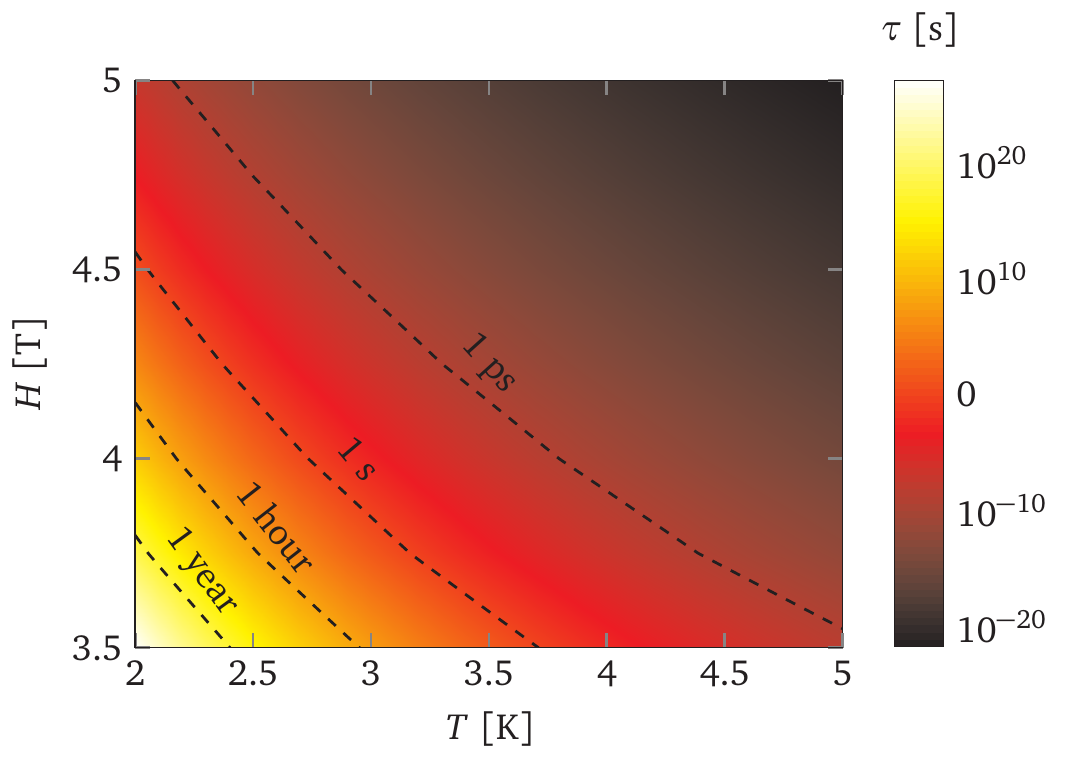}
	\caption
	{
	    Lifetime $\tau$ of an isolated skyrmion in a periodic two-dimensional system, with $J=1$~meV and $D=0.6$~meV, as a function of temperature $T$ and external magnetic field $H$.
	    The lifetime is given on a logarithmic scale with isolines ranging from $1$~ps up to $1$~year.
	    Due to the fact that only a single transition mechanism is taken into account, the structure of the graph is simple.
    }
	\label{fig: htst sk collapse}
\end{figure}
The implementation has been verified against \mbox{\textit{UppASD}},~\cite{skubic_method_2008} and we additionally present an example for the calculation of the lifetime $\tau$ of an isolated skyrmion in a two-dimensional system.
As parameters we chose $J=1$ and $D=0.6~J$ and only the radial collapse mechanism is considered, making for a simple structure of the dependence on external field and temperature.
Note that this example is purely illustrative and while larger skyrmions would exhibit longer lifetimes, the parameters are chosen to produce a small skyrmion in order to reduce the computational effort.
Fig.~\ref{fig: htst sk collapse} shows the results for an external field varied between $3.5$~T and $5$~T and temperature between $2$~K and $5$~K.
HTST as well as Langers theory,~\cite{langer_statistical_1969} which is closely related, have recently both been used to calculate skyrmion lifetimes,~\cite{bessarab_lifetime_2018,desplat_thermal_2018,von_malottki_skyrmion_2018} showing that energy barriers are in general not enough to estimate the stability of metastable magnetic states.
There are two translational zero modes at the initial state minimum, while -- due to the lattice discretisation and the defect-like shape of the skyrmion at the saddle point -- there are no zero modes at the saddle point.
Consequently, the transition rate prefactor has a linear temperature dependence.
%
%
\subsection{Minimum Mode Following Method}
To find the first order saddle points on the energy surface, without prior knowledge of the possible final states, the minimum mode following method~\cite{mueller_duplication_2018} can be used.
The effective force acting on a spin configuration is defined as
\begin{equation}
	F^\mathrm{eff} = F - 2 (F\cdot{\hat\lambda}) {\hat\lambda}\; ,
\label{eq: effective force}
\end{equation}
where $F=-\nabla \mathcal{H}$ is the negative gradient of the energy and $\hat\lambda$ is the normalized eigenvector corresponding to the lowest, negative eigenvalue of the Hessian matrix of second derivatives.
Note that these vectors and the dot product are $3N$-dimensional for a system with $N$ spins.
The calculation of second derivatives requires further attention, as the requirement of constant length effectively constrains the spins to a sub-manifold $\mathcal{M}^\mathrm{phys} \subset \mathcal{E}$ of an embedding space $\mathcal{E} = \mathbb{R}^{3N}$.
As is shown in Ref.~\onlinecite{mueller_duplication_2018}, the covariant second derivatives, valid at all points of the phase space, can be calculated using a projector-based approach.~\cite{absil_extrinsic_2013}
The corresponding ${2N\times2N}$ Hessian matrix can be represented as
\begin{equation}
	H_{ij} = T_i^T \bar{H}_{ij} T_j - T_i^T I (x^j\cdot\nabla^j\bar{\mathcal{H}}) T_j\; ,
    \label{eq: hessian final}
\end{equation}
where $i$ and $j$ are spin indices, $\bar{\mathcal{H}}$ is the smooth continuation of the Hamiltonian to the embedding space, $\bar{H}_{ij} = \partial^2 \bar{\mathcal{H}}$, $I$ is the ${3\times3}$ unit matrix and $T_i$ is a ${3\times2}$ matrix that transforms into a tangent space basis of spin $i$.
As the Hessian matrix \eqref{eq: hessian final} is represented in the $2N$-dimensional tangent basis, the evaluation of an eigenmode in the $3N$-representation of the embedding space $\mathcal{E}$ requires a transformation back, \textit{i.e.}\ $\lambda|^{3N} = T\lambda|^{2N}$.
Further details on the above mathematical concepts and notations can be found in Appendix~\ref{app: manifold}.
\begin{figure}[tb]
    \centering

    \begin{tikzpicture}
    	\node (img1) at (0,0) {
			 \includegraphics[width=0.9\linewidth]{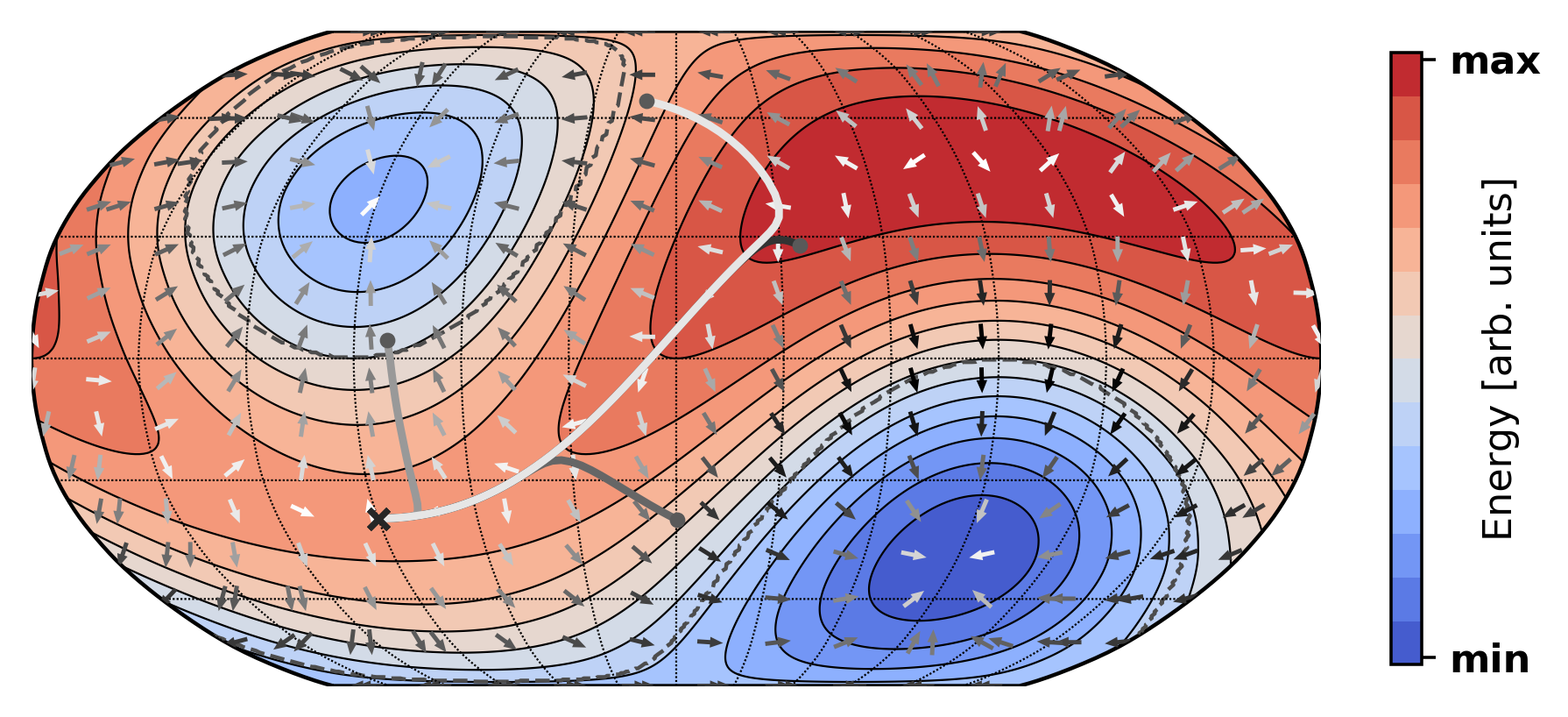} };
    	\node (img2) [below = -0.3 of img1] {
			\includegraphics[width=0.9\linewidth]{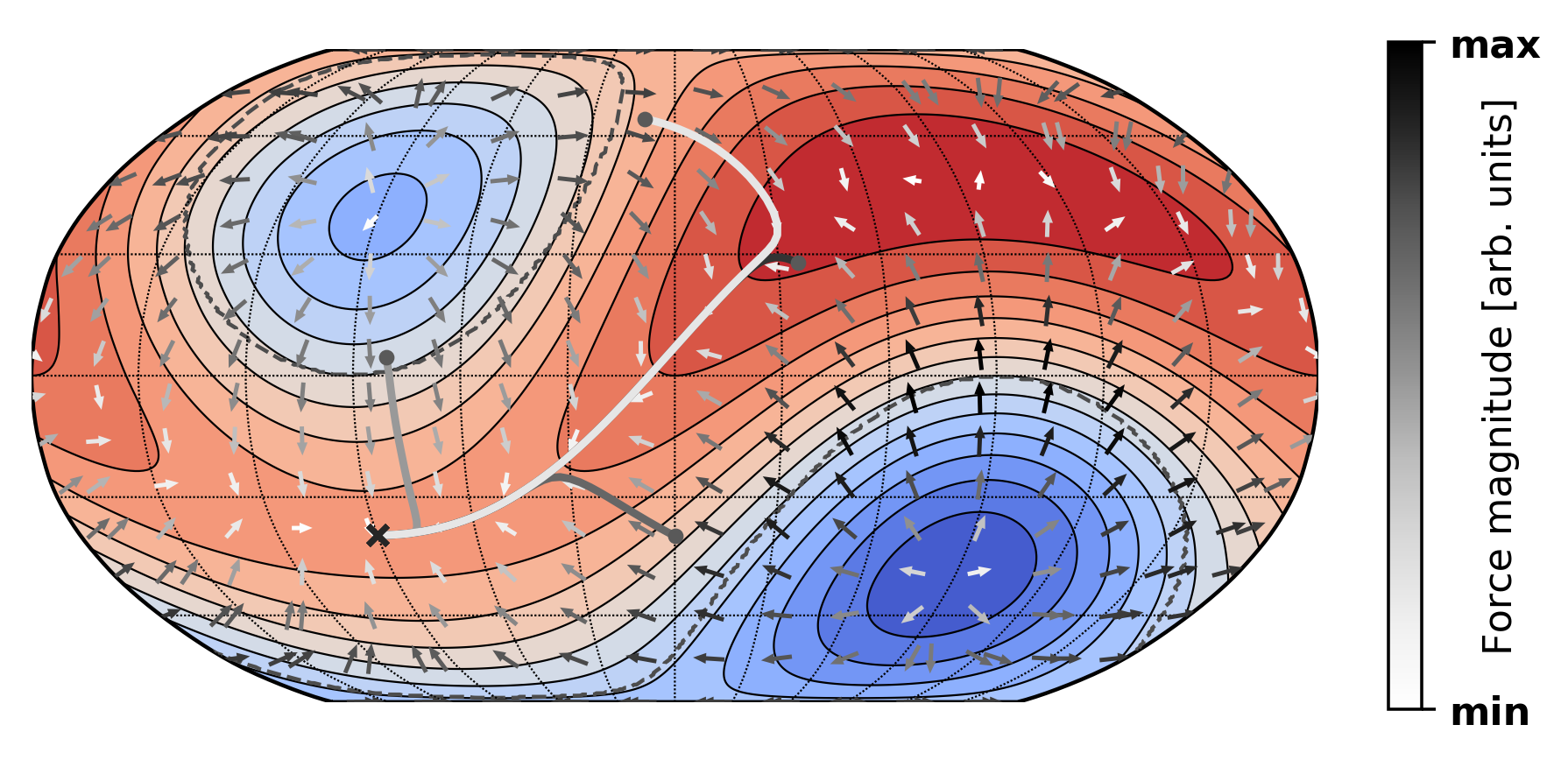} };
		\node [above left=-0.7 and -0.7 of img1] { \small a) };
		\node [above left=-0.7 and -0.7 of img2] { \small b) };
    \end{tikzpicture}

	\caption
	{
        A single spin under the exchange and DMI interaction with another spin.
        The energy landscape is two-dimensional and is projected onto a sphere.
        a) the gradient force field, pointing away from the maximum and towards the minima.
        b) the effective force field, pointing towards the saddle point.
        The resulting paths for four different starting points are shown (black, gray and white lines).
        See Appendix~\ref{app: mode visualization} for a visualization of the minimum mode directions.
    }
	\label{fig: mf single spin}
\end{figure}
For a single spin, the energy landscape and force vectors can be visualized easily as the phase space is two-dimensional.
An illustration of the method is shown in Fig.~\ref{fig: mf single spin} for a system consisting of one movable spin interacting with a second, pinned spin.
The parameters of the Hamiltonian are, relative to the exchange constant $J$,
\begin{equation}
    K=4J\; , \quad \vec{D}=(0,0,1J)\; ,
\end{equation}
where the anisotropy $K$ is used to reduce the symmetry of the energy landscape.
%
The figure illustrates how the minimum mode can be used to invert the right part of the gradient force in order to obtain a force that directs the
system to a first order saddle point. 
The test of a larger and far more complex system has been given in Ref.~\onlinecite{mueller_duplication_2018}, where the minimum mode following method revealed the existence of a skyrmion duplication mechanism.
By defining the force field in the above way, previously unknown transition mechanisms can be found and subsequently used in the calculation of lifetimes.
Applying this saddle point search method to three-dimensional systems will likely identify an even larger variety of mechanisms, as the additional dimension can significantly increase the amount of possible transitions.
%

\section{Conclusions}

The functionality of a comprehensive simulation framework, \textit{Spirit}, for studies of atomic scale magnetic systems is presented and various example applications described.
It is an open source software written in the C$++$ programming language and is available for free under the so-called MIT license (see Ref.~\onlinecite{spirit}).
\textit{Spirit} is a very flexible, high-performance, and interactive tool, able to simulate for example ferromagnets, antiferromagnets, synthetic antiferromagnets, ferrimagnets, noncollinear magnetic structures, vortices or skyrmions.
Arbitrary geometries and interactions can be described, such as bulk systems, thin films, exchange bias, multilayers, nanotubes or core-shell nanoparticles.
The computational domain can be treated by open and periodic boundary conditions and can be subjected to external magnetic fields, temperature and spin-current induced torques.
Due to the fact that it can be used with the Python programming language, \textit{Spirit} can integrate perfectly into  multiscale simulations and workflow automation frameworks, such as ASE~\cite{ASE} or AiiDA.~\cite{AiiDA}
It can be used on most common architectures, such as desktop and laptop computers, clusters or supercomputers and even current day mobile devices.
The calculations can be parallelized both on CPUs and GPUs.

Various simulation methods have been implemented, including Monte Carlo, Landau-Lifshitz-Gilbert dynamics, Langevin dynamics, geodesic nudged elastic band and minimum mode following methods as well as the calculations of transition  rates and lifetimes within the harmonic approximation to transition-state theory.
The basic algorithms of these methods have been outlined, their implementation verified and applications to several systems, such as vortices, domain walls, skyrmions and boobers are described.
The parameters of the simulation can be set and modified in real time through a graphical user interface and the output of the simulations can be visualized easily.

We note that a micromagnetic description of the energetics could easily be implemented in \textit{Spirit} and the micromagnetic calculations
would then be able to make use of the various simulation methods and visualization features.


\begin{acknowledgments}
    The authors acknowledge helpful discussions with Pavel Bessarab, Stephan von Malottki, Jan M\"uller, Jonathan Chico, Filipp N. Rybakov, and Florian Rhiem.
	G.P.M.\ acknowledges funding by the Icelandic Research Fund (grants 185405-051 \& 184949-051) and M.H.\ and S.B.\ from MAGicSky Horizon 2020 European Research FET Open project (\#665095) and from the DARPA TEE program through grant MIPR (\#HR0011831554) from DOI.
\end{acknowledgments}

\appendix


\section{Determination of topological charge for spin density on a lattice}
\label{app: topocharge}
\begin{figure}
    \centering
    \includegraphics[width=6.0cm]{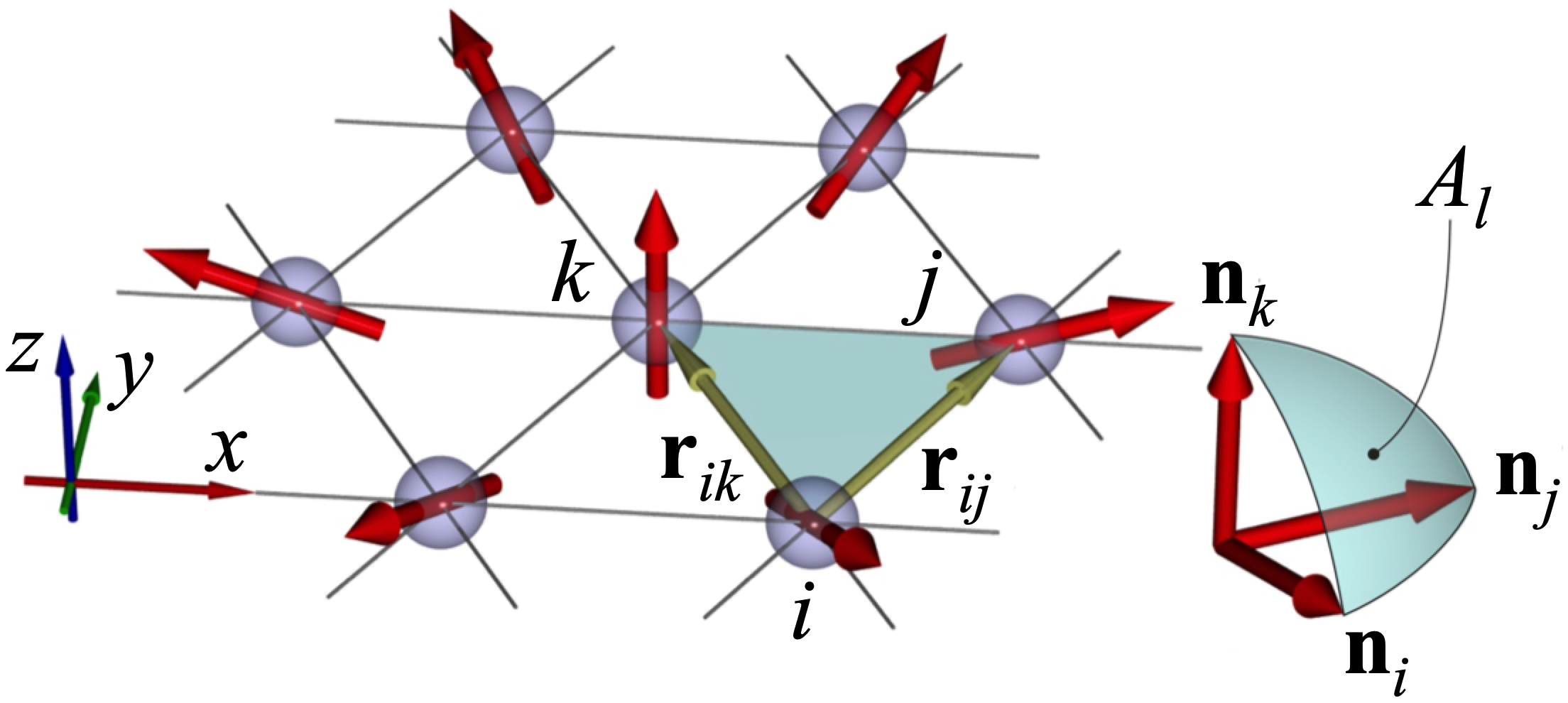}
    \caption{
        Fragment of hexagonal lattice of magnetic spins, which illustrates the definition
        of the topological charge on a discrete lattice as given in the main text.
        $A_l$  is the  area of a spherical triangle defined by vectors $\vec{n}_i$, $\vec{n}_j$, $\vec{n}_k$
        located at the vertices of a triangle of lattice points (indicated shaded).
    }
\label{top_charge}
\end{figure}

For the  proper definition of the topological charge of a discrete lattice of spins $\vec{n}(x_i,y_i)$, where $i$ runs over all lattice sites, we follow the definition given by Berg and L\"uscher,\cite{Berg81} and arrive at the following expression:
\begin{equation}
    Q= \frac{1}{4\pi}\sum_l A_l,
\end{equation}
with
\begin{equation}
    \cos\left(\frac{A_l}{2}\right)=\frac{1  + \vec{n}_i \cdot \vec{n}_j + \vec{n}_i \cdot \vec{n}_k + \vec{n}_j \cdot \vec{n}_k}
    {\sqrt{2\left(1+\vec{n}_i\vec{n}_j\right)\left(1+\vec{n}_j\vec{n}_k
    \right)\left(1+\vec{n}_k\vec{n}_i\right)}}
\label{A_l}
\end{equation}
where $l$ runs over all elementary triangles of the hexagonal lattice, and $A_l$ is the solid angle, \textit{i.e.} the area of the spherical triangle with vertices $\vec{n}_i$, $\vec{n}_j$, $\vec{n}_k$, see Fig.~\ref{top_charge}.
The sign of $A_l$  is determined as
$\operatorname{sign}\left(A_l\right)= \operatorname{sign}\left[\vec{n}_i\cdot\left(\vec{n}_j\times\vec{n}_k\right)\right]$.

The sites $i$, $j$, $k$  of each elementary triangle are numbered in a counter-clockwise sense relative to the surface normal vector $\nvec{r}_\perp$ pointing in positive direction of the $z$-axis.
The latter means that the numbering should satisfy the condition $\nvec{r}_\perp\cdot(\vec{r}_{ij}\times \vec{r}_{ik})>0$, where $\vec{r}_{ij}$ is a connection vector directed from lattice site $i$ to $j$.

The parameter $A_l$ can be thought of as local topological charge, which takes values in the range of $-2\pi<A_l<+2\pi$.
According to Berg and L\"uscher,\cite{Berg81} there is a set of exceptional spin configurations for which $Q$ is not defined but still measurable as $A_l$ in \eqref{A_l} is defined for all possible spin configuration.
The exceptional spin configurations correspond to the case when a spherical triangle degenerates to a great circle $A_l=2\pi$. In this case the orientation of $A_l$ becomes ambiguous and the position of these elementary triangles $l^*$ are considered as exceptional configurations or topological defects of a two-dimensional magnetic structure.

These topological defects satisfy the following condition:
\begin{equation}
 \vec{n}_i \cdot \left( \vec{n}_j \times \vec{n}_k \right)=0,\quad\textrm{and} 
\quad \left|\vec{n}_i + \vec{n}_j+ \vec{n}_k\right| \leq 1.
\label{top_cond+}
\end{equation}
The elementary triangle $l^*$, for which the condition~\eqref{top_cond+} is satisfied, can  be considered as the position at which the localization of a  topological defect takes place.
It is important to note that the definition of the topological charge given above remains correct only for spatially extended two-dimensional systems.
This means that a topological analysis of the spin structure on a finite size domain is only defined if periodical boundary conditions are present.
In the case of open boundary conditions, strictly speaking, the topological charge is not defined.

\section{Heun's solver}
\label{app: heun depondt}

To simplify the following discussion, we write the LLG equation~\eqref{eq: llg} as
\begin{equation}
    \frac{\partial \vec{n}_i (t)}{\partial t} = \vec{n}_i (t) \times \vec{A}_i \left( t, \{\vec{n}_j(t)\} \right)\; ,
\label{eq: llg supp}
\end{equation}
where $\{\vec{n}_j\}$ is the set of all spins and we keep the explicit time-dependence of $\vec{A}_i$, as the Hamiltonian can be time-dependent, for example when an AC magnetic field is used.
%
%
Heun's method is a common and illustrative way to solve ordinary differential equations (ODEs) by first calculating an intermediate prediction step and then ``averaging'' to obtain the final approximation.
Denoting the time step $\delta t$, for an ODE of the form
\begin{equation}
    \frac{\partial y(t)}{\partial t} = f(t, y(t)) \; ,\quad y(t_0) = y_0 \; ,
\end{equation}
the predicted value $y^p$ is first calculated as
\begin{equation}
    y^p(t+\delta t) = y(t) + \delta t f(t, y(t))
\label{eq: heun predictor}
\end{equation}
and then the approximation for the next step as
\begin{equation}
    \begin{alignedat}{2}
    y(t+\delta t)
        =\ &y(t) \\
        +\ &\delta t \frac{ f(t, y(t)) + f(t+\delta t, y^p(t+\delta t)) }{2} \; .
    \end{alignedat}
\label{eq: heun corrector}
\end{equation}
When applied to the LLG equation, where $f \widehat{=} \vec{n}\times\vec{A}$, this integration scheme obviously does not intrinsically preserve the spin length, requiring the re-normalization of the vectors $\vec{n}_i$ after a given number of iterations, depending on the required precision.
Note that Heun's method falls into the category of Runge-Kutta methods, which function analogously and therefore all have this property.

In order to improve on this, Ref.~\onlinecite{depondt_spin_2009} proposes to make use of the fact that the spins are only allowed to rotate, by writing an appropriate rotation matrix $R_i$, which is calculated directly from the field $\vec{A}_i$.
Applied to Heun's method, the prediction step~\eqref{eq: heun predictor} reads
\begin{equation}
    \vec{n}_i^{\,p}(t+\delta t) = R_i\left( \vec{A}_i(t, \{\vec{n}_j(t)\}) \right) \; \vec{n}_i(t)\; .
\end{equation}
To perform the correction step~\eqref{eq: heun corrector}, one needs the correction field $\vec{A}^c$, which is calculated from the average of the initial and predicted fields:
\begin{equation}
    \vec{A}_i^c = \frac{\vec{A}_i(t, \{\vec{n}_j(t)\}) + \vec{A}_i^p(t+\delta t, \{\vec{n}^p_j(t+\delta t)\})}{2}  
\end{equation}
From this, in turn, the rotation matrix for the correction step $R_i^c \left( \vec{A}_i^c \right)$ is obtained and the final step of the scheme reads
\begin{equation} \label{eq:rotation}
    \vec{n}_i(t+\delta t) = R^c_i \left( \vec{A}_i^c \right) \; \vec{n}_i(t) \; .
\end{equation}
Higher order Runge-Kutta schemes could apply this approach analogously.
%

\section{Semi-implicit midpoint solver}
\label{app: sib}

Instead of using a Runge-Kutta type scheme, as described in Appendix~\ref{app: heun depondt}, Ref.~\onlinecite{mentink_stable_2010} takes a different approach, using the implicit midpoint (IMP) structure to preserve the spin length.
The corresponding prediction step in its implicit form reads
\begin{equation}
    \begin{alignedat}{2}
    \vec{n}_i^p(t+\delta t)
        = \vec{n}_i(t)
        +& \delta t \dfrac{\vec{n}_i(t)+\vec{n}_i^p(t+\delta t)}{2} \times \\
         & \vec{A}_i(t, \{\vec{n}_j(t)\}) \;,
    \end{alignedat}
\label{Eq:predictor}
\end{equation}
from which the corrector step can be analogously calculated as
\begin{equation}
    \begin{alignedat}{2}
    \vec{n}_i(t+\delta t)
        =& \vec{n}_i(t)
        + \delta t \dfrac{\vec{n}_i(t)+\vec{n}_i(t+\delta t)}{2}\times \\
        & \vec{A}_i\left(t+\frac{\delta t}{2}, \left\{\dfrac{\vec{n}_j(t)+\vec{n}_j^p(t+\delta t)}{2}\right\}\right) \; .
    \end{alignedat}
\label{eq:sibf}
\end{equation}
The solutions to these equations can be obtained analytically by rewriting them in a skew matrix form and applying Cramer's rule (see Appendix~\ref{app: sib}).

The implicit midpoint method, which the SIB method bases on, solves differential equations of the form $y'(t) = f(t, y(t))$, \, $y(t_0) = y_0$ (see Eq.~(8) of the main text) and an iteration step is defined as
\begin{equation}
        \begin{alignedat}{2}
        y(t+\delta t) = y(t) + \delta t f\left(t+\frac{\delta t}{2},\dfrac{y(t)+y(t+\delta t)}{2}\right)
        \end{alignedat}\; .
\end{equation}
For the LLG equation~\eqref{eq: llg supp} and a time step $\delta t$ this leads us to
\begin{equation}
    \begin{alignedat}{2}
    \vec{n}_i(t+\delta t)
        = &\vec{n}_i(t)
        + \delta t \dfrac{\vec{n}_i(t) + \vec{n}_i(t+\delta t)}{2}
        \times \\
        &\vec{A}_i\left( t+\frac{\delta t}{2}, \left\{ \dfrac{\vec{n}_j(t) + \vec{n}_j(t+\delta t)}{2} \right\} \right) \,.
    \end{alignedat}
\label{eq: LLG-SIB supp}
\end{equation}
The semi-implicit scheme B (SIB) \cite{mentink_stable_2010} uses a predictor $\vec{n}^p_i$ to reduce the implicitness of the equation above by removing the dependence of $\vec{A}_i$ on $\vec{n}_j(t+\delta t)$.
To preserve the spin length the predictor is obtained with the IMP structure.
\begin{equation}
    \begin{alignedat}{2}
        \vec{n}^p_i(t+\delta t) = \vec{n}_i(t) + &\delta t \dfrac{\vec{n}_i(t)+\vec{n}^p_i(t+\delta t)}{2} \times \\
        &\vec{A}(t, \{\vec{n}_j(t)\}) \;.
    \end{alignedat}
\label{Eq: predictor supp}
\end{equation}
Eq.~\eqref{Eq: predictor supp} can be rewritten as:
\begin{equation}
        \begin{alignedat}{2}
        M \cdot \vec{n}^p(t+\delta t) = M^T \cdot \vec{n}(t)
        \label{Eq:cramers}
        \end{alignedat}
\end{equation}
with the matrix
\begin{equation}
    \begin{alignedat}{2}
    M = I + \mathrm{skew}(\vec{A}) =
            \begin{pmatrix}
                1 & -A_z & A_y\\
                A_z & 1 & -A_x\\
                -A_y & A_x & 1
            \end{pmatrix} \;.
    \end{alignedat}
\end{equation}
The right hand side of Eq.~\eqref{Eq:cramers} can be easily calculated as:
\begin{equation}
    \begin{alignedat}{2}
        M^T \vec{n}_i = \vec{n}_i + \vec{n}_i \times \vec{A}_i =: \vec{a} \;.
    \end{alignedat}
\end{equation}
To solve Eq.~\eqref{Eq:cramers} we use Cramer's rule. The components $n^p_{i,\alpha}$ with $\alpha = x,y,z$ of $\vec{n}^p_i$ are calculated with
\begin{equation}
    \begin{alignedat}{2}
        n^p_{i,\alpha} = \dfrac{\mathrm{det}(\vec{M}^{\alpha})}{\mathrm{det}(\vec{M})}
    \end{alignedat}
\label{eq: sib solution}
\end{equation}
where $M^\alpha$ is the same matrix as $M$ but column $\alpha$ is replaced with the vector $\vec{a}$, for example
\begin{equation}
    \begin{alignedat}{2}
        M^x =
            \begin{pmatrix}
                a_x & -A_z & A_y\\
                a_y & 1 & -A_x\\
                a_z & A_x & 1
            \end{pmatrix} \;.
    \end{alignedat}
\end{equation}
We now use the predictor $\vec{n}^p_i$ in the IMP step~\eqref{eq: LLG-SIB supp} to calculate $\vec{n}_i(t+\delta t)$:
\begin{equation}
    \begin{alignedat}{2}
    \vec{n}_i(t+\delta t)
        &= \vec{n}_i(t)
        + \delta t \dfrac{\vec{n}_i(t)+\vec{n}_i(t+\delta t)}{2} \times \\
        &\vec{A}_i\left(t+\frac{\delta t}{2}, \left\{\dfrac{\vec{n}_j(t)+\vec{n}_j^p(t+\delta t)}{2}\right\}\right) \,.
    \end{alignedat}
\label{eq: sibf supp}
\end{equation}
The correction step is analogous to the prediction step (compare eqs.~\eqref{Eq: predictor supp} and \eqref{eq: sibf supp}), meaning that the scheme~\eqref{eq: sib solution} can be applied to obtain $\vec{n}_i(t+\delta t)$, too. 
%

\section{Velocity projection solver}
\label{app: vp}
This description is derived from Ref.~\onlinecite{bessarab_method_2015}.
Verlet-like methods generally find application in solving second order differential equations of the form $\ddot{x}(t) = F(t, x(t))$, $x(t_0) = x_0$, $\dot{x}(t_0) = v_0$, such as Newtons equation of motion.
One formulation of this method is to increment both the position and the velocity at each time step
\begin{equation}
	x(t+\delta t) = x(t) + \delta t \, v(t) + \frac{1}{2m} \delta t^2 \, F(t)
\end{equation}
\begin{equation}
	v(t+\delta t) = v(t) + \frac{1}{2m} \delta t (F(t) + F(t+\delta t)) \; .
\end{equation}
%
The velocity projection is used to accelerate convergence towards local minima and to avoid overstepping due to momentum.
The velocity at each time step is damped by projecting it on the force
\begin{equation}
	v \to \begin{cases}
    	(v\cdot F)F / |F|^2,\ & (v\cdot F) > 0 \\
    	0 					  & \text{else}
	\end{cases}
\end{equation}
Note that the dot product and norm in this equation denote those of $3N$-dimensional vectors.
To apply this scheme to the energy minimization of a spin system, we therefore no longer solve the LLG equation, but instead pretend that the spins are massive particles moving on the surfaces of spheres.
The force is then simply
\begin{equation}
    \vec{F}_i = - \frac{\partial \mathcal{H}}{\partial \vec{n}_i}\; .
\end{equation}
As the method does not conserve the length of the spins, they should be renormalized after each iteration
\begin{equation}
	\vec{n}_i(t+\delta t) \to \frac{\vec{n}_i(t+\delta t)}{|\vec{n}_i(t+\delta t)|}
\end{equation}
Note that this scheme, too, would most likely benefit from the usage of rotations instead of displacements.

\section{Stochastic LLG}
\label{app: langevin}
\begin{figure}[ht]
\centering
	\includegraphics[width=.9\linewidth]{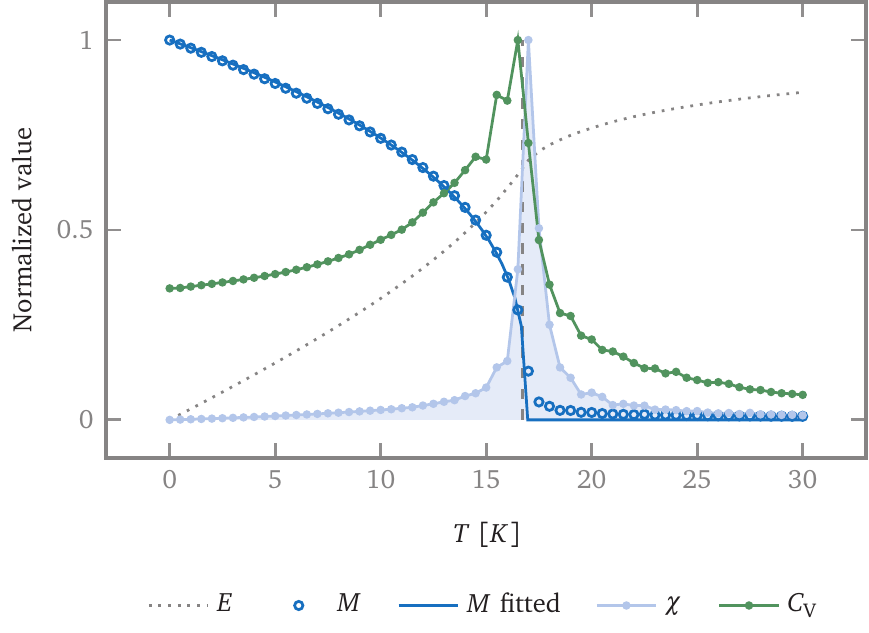}
	\caption{
	    A ${30\times 30\times 30}$ ferromagnet with $J=1$~meV, with an expected critical temperature of $T_\mathrm{C} \approx 16.71$~K.
        The energy per spin $E$ and normalized values of the total magnetization $M$, susceptibility $\chi$, specific heat $c_\mathrm{V}$ and $4^\mathrm{th}$ order Binder cumulant $U_4$ are shown.
        The value obtained from the simulation is $T_c \approx 16.92$~K -- an agreement with expectation of $1.2\%$. The exponent is fitted with $b\approx0.33$.
		At each temperature, $200$k thermalisation steps were made before taking $1$M samples.
    }
	\label{fig: stochastic llg}
\end{figure}

Instead of Monte Carlo, one can also sample the stochastic LLG equation over time.
We present here the results of such sampling for the same system and parameters, as the example shown in Fig.~\ref{fig: monte carlo}. 
Recall the expected critical temperature $T_c = 1.44~J/k_\mathrm{B} \approx 16.71$~K.
Fig.~\ref{fig: stochastic llg} shows the results.

The results shown in Fig.~\ref{fig: stochastic llg} demonstrate the validity of the implementation, as the expected critical temperature of $T_\mathrm{C} \approx 16.71$~K is matched with an error of only $1\%$.
Note, however, the higher number of samples (compared to Monte Carlo) required to obtain this result: at each temperature $200$k thermalisation steps were made before taking $1$M samples.

\section{GNEB tangents and forces}
\label{app: gneb tangents}
\begin{figure}[htb]
	\includegraphics[width=0.5\linewidth]{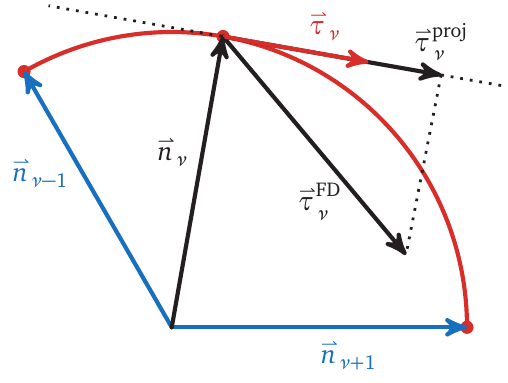}
	\caption
	{
        Schematic visualization of the projection of the tangents for a single-spin system.
		After a tangent $\tau_\nu^\mathrm{FD}$ is determined by finite difference calculation, it needs to be projected onto the tangent plane to the spin configuration so that it correctly points along the path.
		This tangent is denoted $\tau_\nu^\mathrm{proj}$ and can be calculated \textit{e.g.}\ by removing the component in the direction of the image, see Eq.~\eqref{eq: tangent projection}.
		Note that the tangent vector $\tau_\nu$ needs to be normalized, which for a multi-spin system needs to be performed in $3N$ dimensions.
    }
	\label{Fig:Schematic_Tangent_Projection}
\end{figure}
For spin systems, special care has to be taken due to the fact that the phase space is curved (the spins are restricted to unit spheres (see also Appendix~\ref{app: manifold})).
The expression for $l_{\nu,\mu}$ should not be the Euclidean distance norm, but the geodesic (here, the great-circle) distance.
Further, the tangents $\tau_\nu$ need to lie in the tangent space to their corresponding image.
One may correct the tangents for example by a simple projection, orthogonalizing the corresponding $3$-component subvectors with respect to the spins
\begin{equation}
	\vec{\tau}_{\nu,i} \to \vec{\tau}_{\nu,i} - (\vec{\tau}_{\nu,i}\cdot \vec{n}_{\nu,i})\vec{n}_{\nu,i}\;.
\label{eq: tangent projection}
\end{equation}
After this, the tangent needs to be re-normalized $\tau_\nu \to \tau_\nu / |\tau_\nu|$.
This tangent projection is illustrated for a single spin in Fig.~\ref{Fig:Schematic_Tangent_Projection}.
As the spring forces are constructed from tangent vectors, they are by definition in the tangent space.
Finally, for the energy gradient force, the same scheme as for the tangents can be applied and we write for each spin
\begin{equation}
	\vec{F}^\mathrm{E}_{\nu,i} \to \vec{F}^\mathrm{E}_{\nu,i}  - (\vec{F}^\mathrm{E}_{\nu,i} \cdot \vec{n}_{\nu,i}) \vec{n}_{\nu,i}\;.
\label{eq: gradient force orthogonalisation}
\end{equation}

\section{GNEB Parameters of the single spin system}
\label{app: gneb gaussian}
The energy surface of the single-spin system, shown in Fig.~7 in the main text to illustrate the geodesic nudged elastic band method
is defined for a single spin as a sum of Gaussians of the form
\begin{equation}
	\mathcal{H} = \sum\limits_i \mathcal{H}_i  = \sum\limits_i a_i \exp\left( -\frac{(1 - \vec{n}\cdot\vec{c}_i)^2}{2\sigma_i^2} \right)\; ,
\end{equation}
with parameters given in Table~\ref{Table: MMF Single Spin}.
%
\begin{table}[htb]
	\caption{Parameters of the Gaussians in the energy surface of the single-spin system shown in Fig.~7 in the main text.
	}
	\pgfplotstabletypeset
	[
		header=true,
		every column/.style={column type={r}},
		columns/s/.style={column name=$\ \ \ \ \ \sigma$},
		columns/cx/.style={column name=$c_x$},
		columns/cy/.style={column name=$c_y$},
		columns/cz/.style={column name=$c_z$},
		every head row/.style={before row=\toprule, after row=\midrule},
		every last row/.style={after row=\bottomrule},
		fixed,
		fixed zerofill,
		precision=2
	]{
	     a      s     cx     cy   cz
        -1.1   0.06  -0.2   0.0  -0.9
         0.8   0.15  -1.0   0.2  -0.2
        -0.9   0.1    1.0  -0.2  -0.1
         0.09  0.03   0.8   0.5  -0.8
         0.15  0.07   0.8  -0.5  -0.7
        -0.9   0.1    0.5   1.2  -0.4
        -0.9   0.1    0.2  -0.9  -0.4
	}
	\label{Table: MMF Single Spin}
\end{table}

\section{Details on the curved manifold}
\label{app: manifold}
The following has been detailed in the supplementary material of Ref.~\onlinecite{mueller_duplication_2018}, but the key ideas are reproduced here.
Both the HTST and MMF methods require the calculation and diagonalization of the Hessian matrix.
However, when treating Riemannian manifolds, the second derivatives do not have an intrinsic geometrical meaning and therefore need to be treated with special care.~\cite{Nakahara2003}
In a spin system where the spin length is fixed, the manifold $\mathcal{M}_\mathrm{phys}$ of physical states is composed of the direct product of $N$ spheres
\begin{equation}
	\mathcal{M}_\mathrm{phys} = \bigotimes\limits_{i=1}^{N} S^2 \subset \mathbb{R}^{3N}\; .
\end{equation}
Hence, $\mathcal{M}_\mathrm{phys}$ is a submanifold of the embedding euclidean space $\mathcal{E} = \mathbb{R}^{3N}$.
It turns out to be convenient to treat the spins and derivatives with respect to their orientations in a $3N$-dimensional cartesian representation.
This also avoids problems of other representations, such as the singularities which arise at the poles of spherical coordinates.
%
The derivatives in the embedding space $\mathcal{E}$ are readily calculated by extending the Hamiltonian $\mathcal{H}$, which is defined on $\mathcal{M}_\mathrm{phys}$ to a function $\bar{\mathcal{H}}$ on $\mathcal{E}$.
While we denote the gradient taken in the embedding space $\mathcal{E}$ as $\partial\bar{\mathcal{H}}$, the gradient taken on the manifold  $\mathcal{M}_\mathrm{phys}$ has to lie in the tangent space to the manifold, which we write as a projection $P_x\partial\bar{\mathcal{H}}$.
The Hessian matrix of second derivatives in the embedding space $\mathcal{E}$ is denoted $\partial^2\bar{\mathcal{H}}$.

In this extrinsic view onto the spin manifold, the covariant second derivatives can be extracted from a projector approach,~\cite{absil_extrinsic_2013} where for any scalar function $f$ on the manifold $\mathcal{M}_{\mathrm{phys}} $, the covariant Hessian is defined as
\begin{equation}
	\mathrm{Hess}~f({x})[{z}]  = P_{{x}} \partial^2 \bar{f}({x}) {z} + W_{{x}} ({{z}}, P_{{x}}^\perp \partial \bar{f})\; .
\label{eq: supplement hessian generalized}
\end{equation}
%
$W_x$ denotes the Weingarten map, which, for a spherical manifold, for any vector $v$ at a point $x$ is given by
\begin{equation}
	W_x(z,v) = -z x^T v\; ,
\end{equation}
where $z$ is a tangent vector to the sphere at $x$.
To calculate the Hessian, we insert $v=P_x^\perp \partial \bar{\mathcal{H}}$ and retrieve
\begin{equation}
\begin{alignedat}{1}
    W_x (z, P_x^\perp \partial \bar{\mathcal{H}})
      &= -z x^T P_x^\perp \partial  \bar{\mathcal{H}}\\
      &= -z x^T x x^T \partial  \bar{\mathcal{H}}\\
      &= -z x^T \partial  \bar{\mathcal{H}}\; ,
\end{alignedat}
\end{equation}
where $x^T \partial \bar{\mathcal{H}}$ is the scalar product of the spin with the gradient. 
To illustrate the implementation in \textit{Spirit},~\cite{spirit} we switch notation to matrix representation and drop the subscript $x$.
For spin indices $i$ and $j$, the gradient $\partial \bar{\mathcal{H}}$ can be written as a $3$-dimensional object $\nabla_i\bar{\mathcal{H}}$ and the second derivative $\partial^2\bar{\mathcal{H}}$ as a matrix $\bar{H}$.
In Euclidean representation, the Hessian of Eq.~\eqref{eq: supplement hessian generalized} becomes as a ${3N\times3N}$ matrix 
\begin{equation}
    H|^{3N} = (H_{ij}|^{3N} ) = 
        \begin{pmatrix}
        H_{11}|^{3N}  & H_{12}|^{3N} & \cdots \\
        H_{21}|^{3N} & H_{22}|^{3N} & \cdots \\
        \vdots & \vdots & \ddots
        \end{pmatrix}
\end{equation}
consisting of $N^2$ blocks, each corresponding to a different spin-spin subspace.
It is obtained by acting with Eq.~\eqref{eq: supplement hessian generalized} on the euclidean basis vectors of the embedding space $\mathcal{E}$.
These subspace matrices of size ${3\times 3}$ are given by
\begin{equation}
	H_{ij}|^{3N} = P_i \bar{H}_{ij} - \delta_{ij} I \vec{n}_j\cdot\nabla_j\bar{\mathcal{H}}\; ,
\end{equation}
where $I$ denotes the ${3\times3}$ unit matrix.
The matrix $H|^{3N}$ of course describes $3N$ degrees of freedom, while there can only be $2N$ physical eigenmodes of the spins, spanning the tangent space to the spin configuration.
In order to remove the unphysical degrees of freedom in the embedding space $\mathcal{E}$, is is sufficient to transform the matrix into a tangent space basis, which we can write as
$H_{ij} = T_i^T H_{ij}|^{3N} T_j$, where $T_i$ is the basis transformation matrix of spin $i$ fulfilling $T^TP = T^T$ and $T^TT = I|^{2N}$.
The true Hessian $H = (H_{ij})$ of Eq.~\eqref{eq: supplement hessian generalized} in the ${2N\times2N}$ matrix representation, containing only the physical degrees of freedom, is therefore defined as
\begin{equation}
	H_{ij} = T_i^T \bar{H}_{ij} T_j - T_i^T I (\vec{n}_j\cdot\nabla_j\bar{\mathcal{H}}) T_j\; ,
\end{equation}
Note that this reduction of dimensionality also improves the numerical efficiency of the diagonalization.
As the eigenmodes $\lambda|^{2N}$ are represented in the tangent basis, the $3N$ representation needs to be calculated by $\lambda|^{3N} = T\lambda|^{2N}$.

%
While the ${3\times2}$ basis matrix $T_i$ can be calculated quite arbitrarily by choice of two orthonormal vectors, tangent to the spin $\vec{n}_i$, we found it convenient to use the unit vectors of spherical coordinates $\theta$ and $\varphi$
\begin{equation}
\begin{alignedat}{1}
	T = \{\vec{e}_\theta, \vec{e}_\varphi\}
      &= \begin{pmatrix}
          \cos\theta\cos\varphi & -\sin\varphi \\           
          \cos\theta\sin\varphi &  \cos\varphi \\
          -\sin\theta & 0
        \end{pmatrix} \\
      &= \begin{pmatrix}
          zx/r_{xy} & -y/r_{xy} \\           
          zy/r_{xy} &  x/r_{xy} \\
           - r_{xy} & 0
        \end{pmatrix}
\end{alignedat}
\end{equation}
where $r_{xy} = \sin\theta = \sqrt{1-z^2}$.
Note that the poles need to be excluded, but since the basis does not need to be continuous over the manifold, one may \textit{e.g.}\ orthogonalize $\vec{e}_x$ and $\vec{e}_y$ with respect to the spin vector to obtain suitable tangent vectors.
%

%
Finally, the Hessian matrix in the embedding space $\mathcal{E} = \mathbb{R}^{3N}$ is needed, denoted $\bar{H}_{ij}|^{3N}$. 
As the atomistic Hamiltonian can generally be written in matrix form
\begin{equation}
	\begin{alignedat}{2}
	\mathcal{H} =
	    - \sum_j^N A_{ij} \vec{n}_j
	    - \sum\limits_{\braket{ij}} \vec{n}_i B_{ij} \vec{n}_j
	\end{alignedat}
\end{equation}
where $A_{ij}$ are matrices of size ${3\times3}$ describing the linear contributions, such as the Zeeman term, and $B_{ij}$ are matrices describing the quadratic contributions, such as anisotropy, Exchange, DMI and dipolar interactions.
The Hessian matrix is then naturally given by
\begin{equation}
    \bar{H}_{ij} =\partial^2\bar{\mathcal{H}} = - 2 B_{ij}\; .
\end{equation}

\section{Minimum modes in the interacting spin system}
\label{app: mode visualization}

The following Fig.~\ref{fig: mf eigenmodes} illustrates how the minimum eigenmode unit vector $\hat{\lambda}$ is oriented and in which direction, therefore, the gradient force is inverted.
Recall Eq.~\eqref{eq: effective force}, which can be written

\begin{equation}
	F^\mathrm{eff} = -\nabla\mathcal{H} + 2 ({\hat\lambda}\cdot\nabla\mathcal{H}) {\hat\lambda}\; ,
\end{equation}

\begin{figure}[ht]
    \centering
	\includegraphics[width=0.9\linewidth]{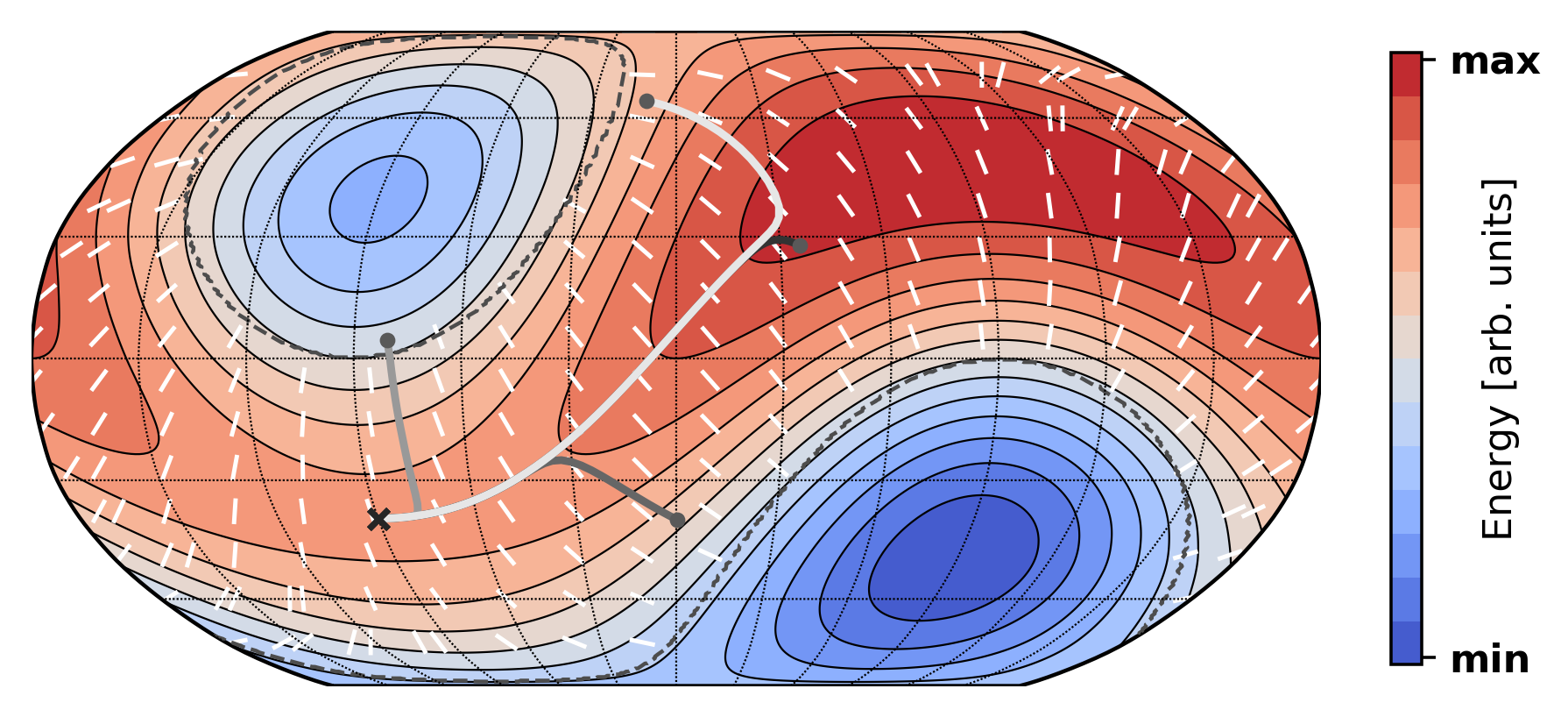}

	\caption
	{
        Field of minimum eigenmodes of a single spin in anisotropy and the interaction field of a second, pinned spin.
        The minimum mode following paths are shown in gray colors.
        The dashed lines show the separation of the convex regions around the minima from the rest of the configuration space.
    }
	\label{fig: mf eigenmodes}
\end{figure}




\begin{thebibliography}{99}


\bibitem{hoffmann_antiskyrmions_2017}{
M. Hoffmann, B. Zimmermann, G. P. M\"uller, D. Sch\"urhoff, N. S. Kiselev, C. Melcher, and S. Bl\"ugel,
"Antiskyrmions Stabilized at Interfaces by Anisotropic {{Dzyaloshinskii}}-{{Moriya}} Interactions,"
Nat. Commun. \textbf{8}, 308 (2017).
}

\bibitem{zutic_spintronics_2004}{
I. \v{Z}uti\'c, J. Fabian, and S. Das Sarma.
"Spintronics: Fundamentals and applications,"
Rev. Mod. Phys. \textbf{76} (2004).
}

\bibitem{bader_colloquium_2006}{
S. D. Bader.
"Colloquium: Opportunities in nanomagnetism,"
Rev. Mod. Phys. \textbf{78} (2006).
}

\bibitem{brown_micromagnetics_1963}{
W. F. Brown,
"Micromagnetics,"
Interscience Publishers (1963).
}

\bibitem{spirit}
{Spirit -- spin simulation framework (see

\url{https://spirit-code.github.io}).
}

\bibitem{oommf}{
M.J. Donahue and D.G. Porter,
"OOMMF User's Guide, Version 1.0,"
Interagency Report NISTIR 6376,
National Institute of Standards and Technology, Gaithersburg, MD (1999).
}

\bibitem{mumax}{
A. Vansteenkiste, J. Leliaert, M. Dvornik, M. Helsen, F. Garcia-Sanchez, and B. Van Waeyenberge,
"The design and verification of MuMax3,"
AIP Advances \textbf{4} 107133 (2014).
}

\bibitem{skubic_method_2008}{
B. Skubic, J. Hellsvik, L. Nordstr\"om, and O. Eriksson,
"A Method for Atomistic Spin Dynamics Simulations: Implementation and Examples,"
J. Phys.: Condens. Matter \textbf{20} 315203 (2008).
}

\bibitem{evans_atomistic_2014}{
R. F. L. Evans, W. J. Fan, P. Chureemart, T. A. Ostler, M. O. Ellis, and R. W. Chantrell,
"Atomistic Spin Model Simulations of Magnetic Nanomaterials,"
J. Phys. Cond. Mat. \textbf{26} 10, 0953-8984 (2014).
}

\bibitem{rybakov_chiral_2018}{
F. N. Rybakov and N. S. Kiselev,
"Chiral Magnetic Skyrmions with Arbitrary Topological Charge ("skyrmionic sacks"),"
arXiv:1806.00782 (2018).
}

\bibitem{nowak_thermally_2001}{
U. Nowak,
"Thermally Activated Reversal in Magnetic Nanostructures,"
Annual Reviews of Computational Physics \textbf{IX}, 105 (2001).
}

\bibitem{bessarab_method_2015}{
P. F. Bessarab, V. M. Uzdin, and H. J\'onsson,
"Method for Finding Mechanism and Activation Energy of Magnetic Transitions, Applied to Skyrmion and Antivortex Annihilation,"
Comp. Phys. Comm. \textbf{196} 335 (2015).
}

\bibitem{mueller_duplication_2018}{
G. P. M\"uller, P. F. Bessarab, S. M. Vlasov, F. R. Lux, N. S. Kiselev, S. Bl\"ugel, V. M. Uzdin, and H. J\'onsson,
"Duplication, Collapse, and Escape of Magnetic Skyrmions Revealed Using a Systematic Saddle Point Search Method,"
Phys. Rev. Lett. \textbf{121} 19 197202 (2018).
}

\bibitem{bessarab_harmonic_2012}{
P. F. Bessarab, V. M. Uzdin, and H. J\'onsson,
"Harmonic Transition-State Theory of Thermal Spin Transitions,"
Phys. Rev. B \textbf{85} 184409 (2012).
}

\bibitem{braun_topological_2012}{
H.-B. Braun,
"Topological effects in nanomagnetism: from superparamagnetism to chiral quantum soliton,"
Advances in Physics \textbf{61} 1 (2012).
}

\bibitem{vfrendering}{
VFRendering -- A vector field rendering library
(see

\url{https://github.com/FlorianRhiem/VFRendering}).
}

\bibitem{liu_binding_2018}{
Y. Liu, R. Lake, and J. Zang,
"Binding a Hopfion in Chiral Magnet Nanodisk,"
arxiv:1806.01682 (2018).
}

\bibitem{zheng_experimental_2018}{
F. Zheng, F. N. Rybakov, A. B. Borisov, D. Song, S. Wang, Z.-A. Li, H. Du, N. S. Kiselev, J. Caron, A. Kov\'acs, M. Tian, Y. Zhang, S. Bl\"ugel, and R. E. Dunin-Borkowski,
"Experimental Observation of Chiral Magnetic Bobbers in B20-Type FeGe,"
Nat. Nanotechnol. \textbf{13}, 451-455 (2018).
}

\bibitem{du_interaction_2018}{
H. Du, X. Zhao, F. N. Rybakov, A. B. Borisov, S. Wang, J. Tang, C. Jin, C. Wang, W. Wei, N. S. Kiselev, Y. Zhang, R. Che, S. Bl\"ugel, and M. Tian,
"Interaction of Individual Skyrmions in a Nanostructured Cubic Chiral Magnet,"
Phys. Rev. Lett. \textbf{120}, 197203 (2018).
}

\bibitem{hagemeister_controlled_2018}{
J. Hagemeister, A. Siemens, L. R\'ozsa, E. Y. Vedmedenko, and R. Wiesendanger,
"Controlled Creation and Stability of $k\ensuremath{\pi}$ Skyrmions on a Discrete Lattice,"
Phys. Rev. B \textbf{97}, 174436 (2018).
}

\bibitem{redies_distinct_2018}{
M. Redies, F. R. Lux, P. M. Buhl, G. P. M\"uller, N. S. Kiselev, S. Bl\"ugel, and Y. Mokrousov,
"Distinct Magnetotransport and Orbital Fingerprints of Chiral Bobbers,"
arXiv:1811.01584 (2018).
}

\bibitem{juspin}{
Web interface for Spirit
(see \url{https://juspin.de}).
}

\bibitem{ASE}{
A. H. Larsen, J. J. Mortensen, J. Blomqvist,
I. E. Castelli, R. Christensen, M. Dułak, J. Friis,
M. N. Groves, B. Hammer, C. Hargus, E. D. Hermes,
P. C. Jennings, P. B. Jensen, J. Kermode, J. R. Kitchin,
E. L. Kolsbjerg, J. Kubal, K. Kaasbjerg,
S. Lysgaard, J. Bergmann Maronsson, T. Maxson, T. Olsen,
L. Pastewka, A. Peterson, C. Rostgaard, J. Schiøtz,
O. Sch\"utt, M. Strange, K. S. Thygesen, T. Vegge,
L. Vilhelmsen, M. Walter, Z. Zeng, and K. W. Jacobsen,
"The Atomic Simulation Environment—A Python library for working with atoms,"
J. Phys.: Condens. Matter \textbf{29} 273002 (2017).
(see \url{https://wiki.fysik.dtu.dk/ase})
}

\bibitem{AiiDA}{
G. Pizzi, A. Cepellotti, R. Sabatini, N. Marzaria, and B. Kozinsky,
"AiiDA: automated interactive infrastructure and database for computational science,"
Comp. Mat. Sci. \textbf{111} 218 (2016).
(see also \url{http://www.aiida.net/})
}

\bibitem{aharoni_introduction_2000}{
A. Aharoni,
"Introduction to the Theory of Ferromagnetism,"
Oxford University Press (2000).
}

\bibitem{rado_magnetism_1963}{
G. T. Rado,
Magnetism: a treatise on modern theory and materials. 3. Spin arrangements and crystal structure, domains, and micromagnetics.
Academic Press (1963).
}

\bibitem{hoffmann_systematic_2018}{
M. Hoffmann and S. Bl\"ugel,
"Systematic derivation of realistic spin-models for beyond-Heisenberg solids from microscopic model,"
arXiv:1803.01315 (2018).
}

\bibitem{szilva_interatomic_2013}{
A. Szilva, M. Costa, A. Bergman, L. Szunyogh, L. Nordström, and O. Eriksson,
"Interatomic Exchange Interactions for Finite-Temperature Magnetism and Nonequilibrium Spin Dynamics,"
Phys. Rev. Lett. \textbf{111} (2013).
}

\bibitem{kroenlein_magnetic_2018}{
A. Krönlein, M. Schmitt, M. Hoffmann, J. Kemmer, N. Seubert, M. Vogt, J. Küspert, M. Böhme, B. Alonazi, J. Kügel, H. A. Albrithen, M. Bode, G. Bihlmayer, and S. Blügel,
"Magnetic Ground State Stabilized by Three-Site Interactions: Fe/Rh(111),"
Phys. Rev. Lett. \textbf{120}, 207202 (2018).
}

\bibitem{heinze_spontaneous_2011}{
S. Heinze, K. von Bergmann, M. Menzel, J. Brede, A. Kubetzka, R. Wiesendanger, G. Bihlmayer, and S. Blügel,
"Spontaneous atomic-scale magnetic skyrmion lattice in two dimensions,"
Nat. Phys. \textbf{7} 713 (2011).
}


\bibitem{hayashi_ddi_1996}{
N. Hayashi, K. Saito, and Y. Nakatani,
"Calculation of Demagnetizing Field Distribution Based on Fast Fourier Transform of Convolution,"
Japanese Journal of Applied Physics \textbf{12} (1996).
}

\bibitem{fftw}{
M. Frigo and S. G. Johnson,
"The Design and Implementation of FFTW3,"
Proceedings of the IEEE 93 (2), 216 (2005).
Invited paper, Special Issue on Program Generation, Optimization, and Platform Adaptation
}

\bibitem{cufft}{
cuFFT https://developer.nvidia.com/cufft
}

\bibitem{hubert_systematic_1999}{
A. Hubert and W. Rave,
"Systematic Analysis of Micromagnetic Switching Processes,"
Phys. Status Solidi B \textbf{211} 2 815-829 (1999).
}

\bibitem{binder_monte_1981}{
K. Binder and D. W. Heermann,
"Monte  Carlo  Simulation in Statistical Physics,"
Springer-Verlag, Berlin, (1997).
}

\bibitem{hinzke_monte_1999}{
D.Hinzke and U.Nowak,
"Monte Carlo simulation of magnetization switching in a Heisenberg model for small ferromagnetic particles,"
Comp. Phys. Comm. \textbf{121} 334 (1999).
}

\bibitem{landau_guide_2009}{
D. P. Landau and K. Binder,
"A guide to Monte Carlo simulations in Statistical Physics,"
Cambridge University Press, New York, NY. (2005).
}

\bibitem{binder_finite_1981}{
K. Binder,
"Finite Size Scaling Analysis of Ising Model Block Distribution Functions,"
Z. Phys. B \textbf{43}, 119 (1981).
}



\bibitem{baker_high-temperature_1967}{
G. A. Baker, Jr., H. E. Gilbert, J. Eve, and G. S. Rushbrooke.
High-temperature expansions for the spin 1/2 Heisenberg model,"
Phys. Rev. \textbf{164}, 800 (1967).
}

\bibitem{rocio_modeling_2011}{
Y. Rocio,
"Modeling of Macroscopic Anisotropies Due to Surface Effects in Magnetic Thin Films and Nanoparticles,"
PhD thesis (2011).
}

\bibitem{swendsen_replica_1986}{
R.~H. Swendsen and J.-S. Wang,
"Replica Monte Carlo Simulation of Spin-Glasses,"
Phys. Rev. Lett. \textbf{57} 2607 (1986).
}

\bibitem{hukushima_exchange_1996}{
K. Hukushima and K. Nemoto,
"Exchange Monte Carlo Method and Application to Spin Glass Simulations,"
J. Phys. Soc. Jap. \textbf{65} 1604 (1996).
}

\bibitem{bottcher_b-t_2017}{
M. B\"ottcher, S. Heinze, S. Egorov, J. Sinova, and B. Dup\'e,
"$B$-$T$ Phase Diagram of Pd/Fe/Ir(111) Computed with Parallel Tempering Monte Carlo,"
arXiv:1707.01708 (2017).
}

\bibitem{landau_on_1935}{
L.~D. Landau and E. M. Lifshitz,
"On the Theory of the Dispersion of Magnetic Permeability in Ferromagnetic Bodies,"
Physik Z. Sowjetunion \textbf{8} 153 (1935).
}

\bibitem{gilbert_phenomenological_2004}{
T.~L. Gilbert,
"A phenomenological theory of damping in ferromagnetic materials,"
IEEE Transactions on Magnetics, \textbf{40} (6), (2004).
}

\bibitem{brown_thermal_1963}{
W.~F. Brown,
"Thermal Fluctuations of a Single-Domain Particle,"
\textit{Physical Review} \textbf{130} 5, 1677–86 (1963).
}

\bibitem{schieback_numerical_2007}{
C. Schieback, M. Kläui, U. Nowak, U. Rüdiger, and P. Nielaba,
"Numerical Investigation of Spin-Torque Using the Heisenberg Model,"
\textit{The European Physical Journal B} \textbf{59} 4, 429–33 (2007).
}

\bibitem{Berg81}{
B. Berg, M. L\"uscher, 
"Definition and statistical distribution of a topological number in the Lattice $O(3)$ $\sigma$-Model*," Nuclear Physics B \textbf{190}[FS3] 412 (1981).
}

\bibitem{mentink_stable_2010}{
J. H. Mentink, M. V. Tretyakov, A. Fasolino, M. I. Katsnelson, and T. Rasing,
"Stable and Fast Semi-Implicit Integration of the Stochastic Landau–Lifshitz Equation,"
J. Phys. Cond. Mat. \textbf{22} 17, 176001 (2010).
}


\bibitem{bauer_thermally_2011}{
D. S. G. Bauer, P. Mavropoulos, S. Lounis, and S. Blügel,
"Thermally activated magnetization reversal in monatomic magnetic chains on surfaces studied by classical atomistic spin-dynamics simulations,"
J. Phys.: Condens. Matter \textbf{23} 394204 (2011).
}

\bibitem{levente_langevin_2014}{
L. R\'ozsa, L. Udvardi, and L. Szunyogh,
"Langevin spin dynamics based on ab initio calculations: numerical schemes and applications,"
J. Phys.: Condens. Matter \textbf{26} 216003 (2014).
}

\bibitem{depondt_spin_2009}{
P. Depondt and F. G. Mertens,
"Spin Dynamics Simulations of Two-Dimensional Clusters with Heisenberg and Dipole-Dipole Interactions,"
J. Phys.: Condens. Matter \textbf{21} 33 (2009).
}

\bibitem{thiaville_domain_2004}{
A. Thiaville, Y. Nakatani, J. Miltat, and N. Vernier,
"Domain Wall Motion by Spin-Polarized Current: A Micromagnetic Study,"
J. Appl. Phys. \textbf{95} 11 (2004).
}

\bibitem{schryer_motion_1974}{
N.~L. Schryer and L.~R. Walker,
"The motion of $180^o$ domain walls in uniform dc magnetic fields,"
J. Appl. Phys. \textbf{45} (12):5406–5421 (1974)
}

\bibitem{henkelman_improved_2000}{
G. Henkelman and H. J\'onsson,
"Improved Tangent Estimate in the Nudged Elastic Band Method for Finding Minimum Energy Paths and Saddle Points,"
J. Chem. Phys. {\bf 113}, 9978 (2000).
}

\bibitem{henkelman_climbing_2000}{
G. Henkelman, G. P. Uberuaga, and H. J\'onsson,
"A climbing image nudged elastic band method for finding saddle points and minimum energy paths,"
J. Chem. Phys. {\bf 113}, 9901 (2000).
}

\bibitem{rybakov_new_2015}{
F. N. Rybakov, A. B. Borisov, S. Bl\"ugel, and N. S. Kiselev,
"New type of particlelike state in chiral magnets,"
Phys. Rev. Lett. \textbf{115} 117201 (2015).
}

\bibitem{wigner_transition_1938}{
E. Wigner,
"The Transition State Method,"
Transactions of the Faraday Society \textbf{34}, 29 (1938).
}

\bibitem{truhlar_current_1996}{
D. G. Truhlar, B. C. Garrett, and S. J. Klippenstein,
"Current Status of Transition-State Theory,"
J. Phys. Chem. \textbf{100} 12771 (1996).
}

\bibitem{langer_statistical_1969}{
J. S. Langer,
"Statistical theory of the decay of metastable states,"
Annals of Physics \textbf{54} 258 (1969).
}

\bibitem{bessarab_lifetime_2018}{
P. F. Bessarab, G. P. M\"uller, I. S. Lobanov, F. N. Rybakov, N. S. Kiselev, Nikolai, H. J\'onsson, V. M. Uzdin, S. Bl\"ugel, L. Bergqvist, and A. Delin,
"Lifetime of Racetrack Skyrmions,"
Sci. Rep. \textbf{8} 3433 (2018).
}

\bibitem{desplat_thermal_2018}{
L. Desplat, D. Suess, J.-V. Kim, and R.~L. Stamps,
"Thermal stability of metastable magnetic skyrmions: Entropic narrowing and significance of internal eigenmodes,"
Phys. Rev. B \textbf{98}, 134407 (2018).
}

\bibitem{von_malottki_skyrmion_2018}{
S. von Malottki, P. F. Bessarab, S. Haldar, A. Delin, and S. Heinze,
"Skyrmion lifetimes in ultrathin films,"
arXiv:1811.12067 (2018).
}

\bibitem{absil_extrinsic_2013}{
P.-A. Absil, R. Mahony, and J. Trumpf,
"An Extrinsic Look at the Riemannian Hessian,"
In {\it Geometric Science of Information} 361-68, Lecture Notes in Computer Science, Springer, Berlin, Heidelberg (2013).
}


\bibitem{Nakahara2003}{
M. Nakahra,
"Geometry, topology and physics,"
CRC Press (2003).
}



\end{thebibliography}
\end{document}